\documentclass[11pt]{article}
\usepackage{graphicx}
\usepackage{times}
\usepackage{amssymb}
\usepackage[numbers,sort,square,comma]{natbib}
\usepackage[vcentering,dvips]{geometry}
\def\gsim { \lower .75ex \hbox{$\sim$} \llap{\raise .27ex \hbox{$>$}} }
\def\lsim { \lower .75ex \hbox{$\sim$} \llap{\raise .27ex \hbox{$<$}} }
\newcommand{\apj}{ApJ}
\newcommand{\apjl}{ApJL}
\newcommand{\apjs}{ApJS}
\newcommand{\aj}{AJ}
\newcommand{\mnras}{MNRAS}
\newcommand{\nat}{Nature}

\newcommand{\araa}{ARA\&A}
\newcommand{\aapr}{A\&ARv}
\newcommand{\aap}{A\&A}

\newcommand{\pasp}{Astronomical Society of the Pacific}

\newcommand{\physrep}{PhysRep}

\begin{document}

\title{Active Galactic Nuclei -- 
the Physics of Individual Sources and the Cosmic History of Formation and Evolution}

\author{
{Henric Krawczynski$^{1}$ and Ezequiel Treister$^{2}$}
\\
\\
$^{1}$ Washington University in St. Louis\\
$^{2}$ Universidad de Concepci\'{o}n, Departamento de Astronom\'{\i}a, \\ Casilla 160-C, Concepci\'{o}n, Chile\\
\\
Correspondence: Henric Krawczynski (krawcz@wuphys.wustl.edu) and \\
Ezequiel Treister (etreiste@astro-udec.cl)\\
}
\date{\today}
\maketitle

\begin{abstract}
In this paper we give a brief review of the astrophysics of active galactic nuclei (AGN).
After a general introduction motivating the study of AGNs, we discuss our present
understanding of the inner workings of the central engines, most likely accreting
black holes with masses between 10$^6$ and 10$^{10} M_{\odot}$.
We highlight recent results concerning the jets (collimated outflows) of AGNs derived 
from X-ray observations (Chandra) of kpc-scale jets and 
$\gamma$-ray  observations of AGNs (Fermi, Cherenkov telescopes) with jets closely 
aligned with the lines of sight (blazars), and discuss the interpretation of these observations.
Subsequently, we summarize our knowledge about the cosmic history of AGN formation 
and evolution. We conclude with a description 
of upcoming observational opportunities.
\end{abstract}

\section{Motivation}
Active galactic nuclei (AGNs) are galaxies that harbor supermassive black holes (SMBHs) of a few million to 
a few billion solar masses. Whereas it seems likely that all galaxies 
contain one or more supermassive black holes \cite{Mago:98,Gult:09}, 
the black holes in  AGNs give rise to spectacular observational consequences because they 
accrete matter and convert the gravitational energy of the accreted matter 
(and possibly also the rotational energy of the black hole) 
into mechanical and electromagnetic energy. 
 A few of the most salient motivations for the study of AGNs are: 
\begin{description} 
\item[AGN Taxonomy:] AGNs are among the brightest extragalactic sources  and account for a large 
fraction of the electromagnetic energy output of the Universe, motivating their  
taxonomy and statistical characterization. 
The study of AGNs in the nearby Universe shows that 
the  diversity of AGNs can be understood as resulting from observing a smaller number of basic 
AGN types from different viewing angles (see Section \ref{core}).      
\item[Accretion Physics:]
AGNs are powered by the accretion of magnetized plasma. Studies of  
AGN accretion flows complement studies of other accretion flows in astrophysics: accretion onto 
protostars and stars, accretion onto compact stellar remnants (neutron stars and 
stellar mass black holes), and the accretion that powers gamma-ray bursts.
One goal of the studies of AGN accretion flows is to provide a physical explanation 
of the different types of AGNs and their states in terms of the nature of 
their accretion flows and environments (see Section \ref{flows}).
\item[Role in Eco-Systems:] AGNs play an important role for galactic and galaxy cluster 
eco-systems, i.e. their mechanical and electromagnetic power  
contributes to the heating of the interstellar and intracluster medium, 
and thus influences the star formation of the host systems (see Section \ref{feed}).
\item[History through Cosmic Time:]
Deep radio, IR, optical and X-ray observations of AGNs have provided us with a wealth of
information about the cosmic history of the formation and growth of supermassive black 
holes and the evolution of AGNs.  
Related areas of research are to clarify the role of AGNs in re-ionizing the 
intergalactic medium, and to explain the correlation between black hole masses 
and the properties of the host galaxy observed in the local Universe (see Sections \ref{ev} \& \ref{trigger}).
\item[Fundamental Physics:] 
On the most fundamental level, AGNs allow
us to test the theory of general relativity (GR). GR's no-hair theorem states that Kerr (and more 
generally Kerr-Newman) solutions are the only stationary, axially symmetric vacuum solutions 
of the Einstein equations with an event horizon. 
Testing if astrophysical black holes are Kerr black holes thus constitutes a powerful 
test of GR in the observationally poorly constrained strong-gravity regime \cite{Psal:08}.
\item[Astroparticle Physics:]
AGNs are astroparticle physics laboratories. A few examples: The TeV $\gamma$-ray emission from
AGNs tells us that they can accelerate particles to $>$TeV energies and AGNs might even be the sources 
of Ultra High Energy Cosmic Rays. The studies of the broadband emission from AGNs allows us to 
perform time resolved studies of the particle acceleration processes. AGN observations can also 
be used to constrain Lorentz Invariance violations \cite[e.g.][]{Abdo:09}, and to set upper and 
lower limits on extragalactic magnetic fields (see the discussion in \cite{Tayl:11,Brod:12}). 

\item[AGNs as Beacons at Cosmological Distances:] 
The emission from AGNs can be used to study the properties of objects, diffuse matter, 
and radiation fields that are located between us and the AGNs. High-resolution spectra
of high-redshift, low-metallicity quasar absorption line systems have been used to constrain the
relative abundance of the light elements produced during the epoch of the Big Bang Nucleosynthesis \cite{PDG:12}.
Measurements of the Gunn-Peterson optical depths of high-redshift quasars constrain the re-ionization 
history of the intergalactic medium \cite{Fan:06}. 
X-Ray Absorption lines constrain the abundance and properties of warm-hot intergalactic medium \cite{Yao:12}.
The study of the GeV and TeV $\gamma$-ray energy spectra of blazars can be used to constrain the
energy spectrum of the infrared and optical Extragalactic Background Light \cite{Meye:12}.
\end{description} 
The discussion in this paper is limited to recent results concerning the properties and inner workings of 
AGNs, and does not cover the research that uses AGNs as probes of the intervening medium. 
In Sect.\ \ref{jets} we discuss observations of AGN jets and their interpretation. In Sect.\ \ref{cores} 
we describe observations of  the AGN cores, as well as models of AGN accretion and jet formation.
We review recent results concerning the cosmic history of black hole accretion in Sect.\ \ref{ev}, 
and trigger of black hole growth in Sect.\ \ref{trigger}. 
We conclude with a brief description of upcoming observational opportunities afforded by present and
upcoming experiments in Sect.\ \ref{outlook}.
\begin{figure}[tb]
\begin{center}
\includegraphics[angle=0.,height=6cm]{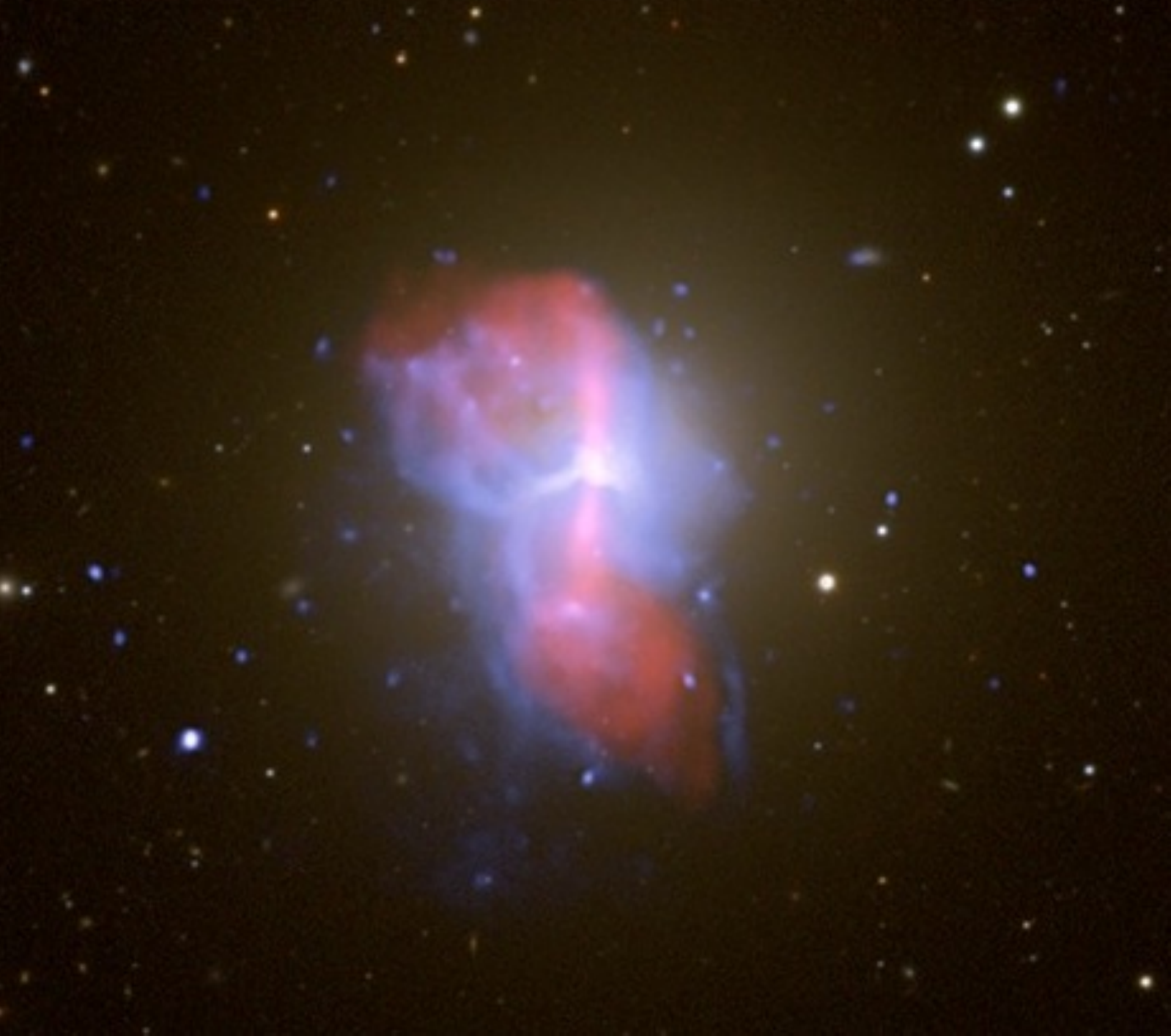}
\includegraphics[angle=0.,height=6cm]{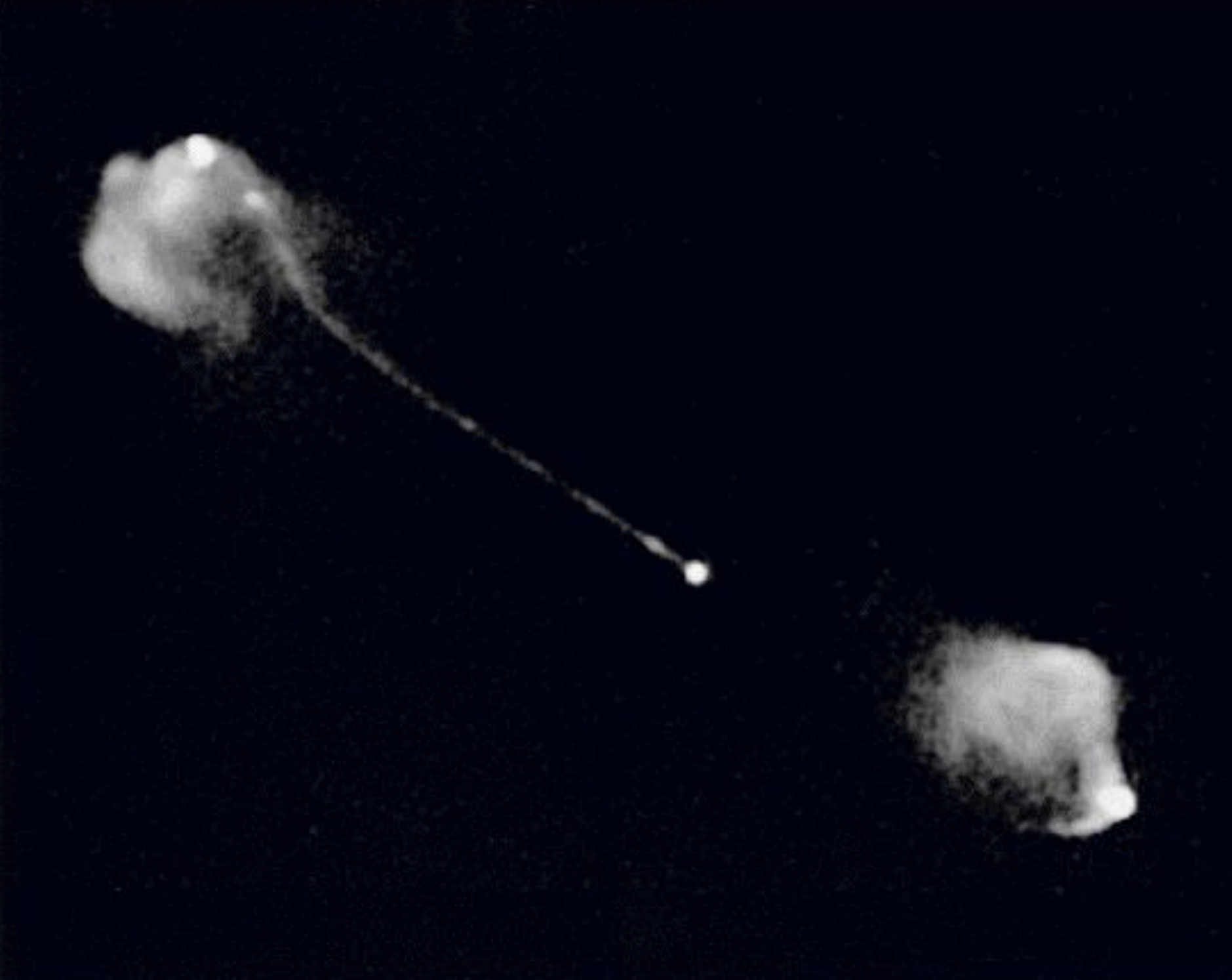}
\caption{Comparison of the morphologies of a FRI radio galaxy (M84, left panel) 
and a FRII radio galaxy (3C 175, right panel). M84 is a massive elliptical galaxy in the Virgo cluster. 
The false-color composite image shows the Very Large Array 4.9~GHz image of the radio galaxy in red \cite{Lian:87}, 
the Chandra 0.5-2~keV image of hot galaxy gas in blue \cite{Fino:08},  and a Sloan Digital Sky Survey optical 
image in yellow and white. The radio galaxy has a projected diameter of $\approx$12 kpc.
The Very Large Array 4.9~GHz image of 3C 175 ($z=0.768$) shows a well-collimated 
jet, and two hot spot complexes typical for FRII radio galaxies.
The projected distance between the two hotspots at the two ends of the radio galaxy is $\approx$370 kpc.
Credits for the left panel: X-ray (NASA/CXC/MPE/A.Finoguenov et al.); 
Radio (NSF/NRAO/VLA/ESO/R.A.Laing et al); Optical (SDSS). 
The right image is a courtesy of NRAO/AUI.
}
\label{FRI}
\end{center}
\end{figure}
 \section{AGN Jets on pc, kpc and Mpc Scales}
 \label{jets}
\subsection{Morphology of Radio Galaxies}
 \label{morph}
Some of the gravitational energy of the material accreted by AGNs is converted into heat and electromagnetic radiation
inside the accretion disk  and is radiated away by the accretion disk. Some of the material processed
through the accretion disk escapes the accretion system as collimated (jets) and uncollimated (winds) outflows.
The event horizon of a non-rotating Schwarzschild black hole is two times the gravitational radius:
\begin{equation}
r_{\rm g}\,=\, \frac{G M_{\rm BH}}{c^2}\,\approx \, 1.48\,\frac{M_{\rm BH}}{10^8 M_{\odot}} \,10^{13}\, {\rm cm}
\end{equation}
Approximately 10\%-20\% of AGNs are radio loud 
(radio to optical spectral index $>$0.35, see e.g. \cite{Ceca:94,Kell:98}),
and show bright extended radio features with sizes up to $\sim$1~Mpc
($3.08\times 10^{24}$ cm). The AGN phenomenon thus spans $\sim$11 orders of magnitudes in size scales.

The first jet of a SMBH was observed in the optical: in 1918, H.\ D.\ Curtis wrote about the object  M~87 in the 
Virgo cluster \cite{Curt:18}:  ``A curious straight ray lies in a gap in the nebulosity in p.a. 20$^{\circ}$, apparently 
connected with the nucleus by a thin line of matter.''  It was not before 1963 that the extragalactic
nature of quasars (quasi-stellar objects, a type of AGNs) was established 
when Schmidt measured the redshift $z=0.158$ of the radio source 3C 273 \cite{Schm:63}.
We now know that radio-loud galaxies come in two qualitatively different types (see Fig.\ \ref{FRI}): 
Fanaroff-Riley Class I (FRI) sources have center-brightened outflows and
Fanaroff-Riley Class II (FRII) exhibit an edge-brightened morphology.
Whereas the separation of the regions of highest radio brightness of FRI sources are 
smaller than  half the size of the radio source, those of FRII sources are larger than  
half the size of the radio source. 
FRI sources are less powerful than FRII sources, with
1.4 GHz luminosities below and above $5\times 10^{25}$ W Hz$^{-1}$, respectively.

\begin{figure}[tb]
\begin{center}
\includegraphics[angle=0.,height=7cm]{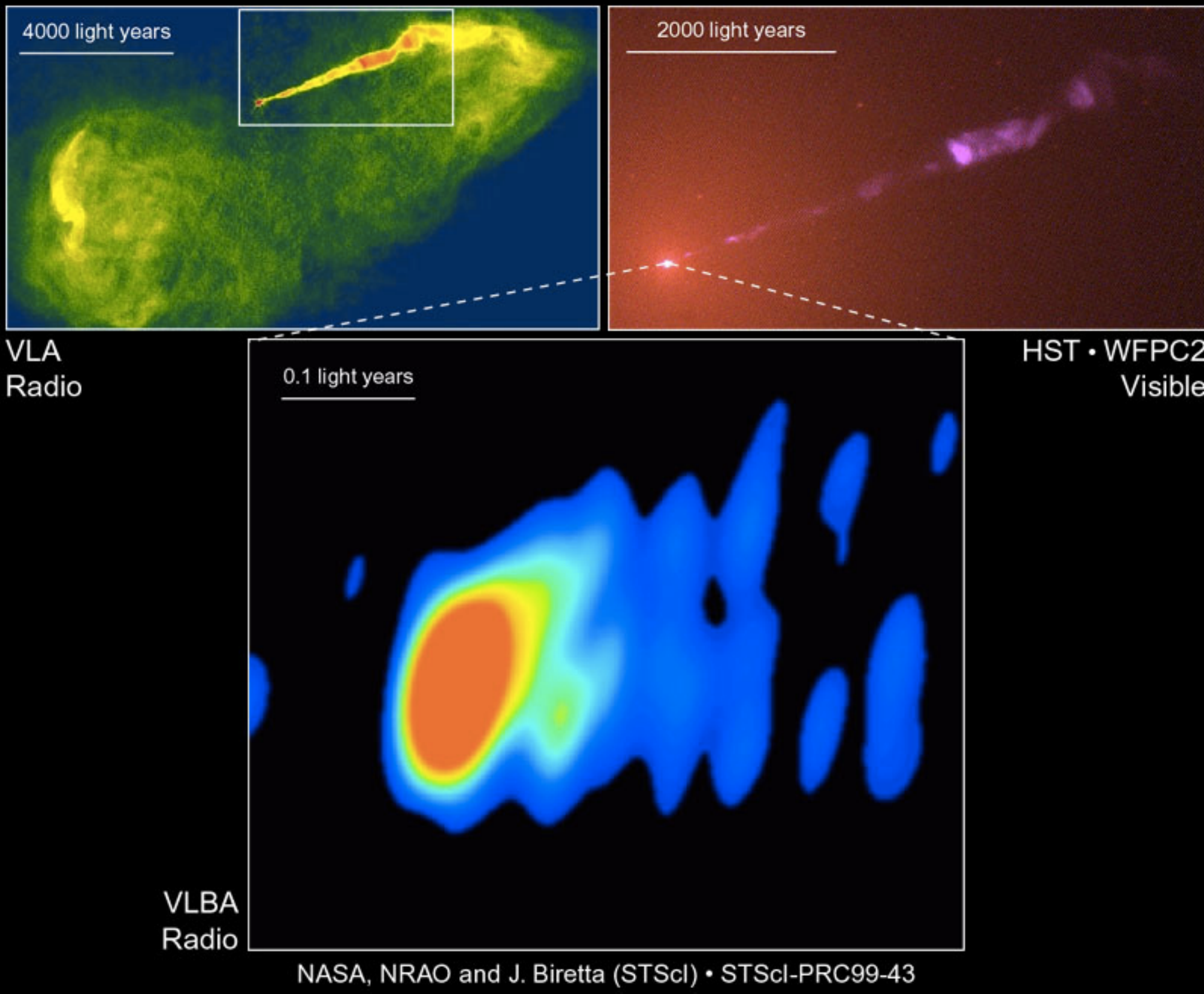}
\caption{Images of the radio galaxy M87 at the center of the Virgo galaxy cluster at different spatial scales and 
in different wavelengths. The Very Large Array radio image (top left) side shows  the kpc-scale jet inflating radio
lobes. The Hubble Space Teelscope optical image  (top right) shows the structure of the kpc-scale jet.
The Very Long Baseline Array image (bottom center)  shows the sub-pc scale jet very close to the black hole.
Credits:  National Radio Astronomy Observatory/National Science Foundation,NASA and John Biretta (STScI/JHU), 
National Radio Astronomy Observatory/Associated Universities, Inc.}
\label{M87}
\end{center}
\end{figure}

The observations of AGNs with single-dish and interferometric radio telescopes and with optical
telescopes have provided us with detailed information about the morphology of radio-loud AGNs. 
As an example, Fig.\ \ref{M87} shows radio and optical images of M~87 with various resolutions, 
zooming in from an image of radio lobes blown by the central engine into the 
intracluster gas to the inner jet imaged with a resolution of a few 10 $r_{\rm g}$.
One distinguishes between the unresolved radio-core (which is likely to coincide with the location
of the SMBH), the well collimated jet (which transports energy away from the SMBH),  
jet knots (locations of increased energy dissipation), hotspots and hotspots complexes 
(where the jet impinges on the ambient medium and most of the mechanical jet energy 
is dissipated in strong shocks),  and lobes of radio-emitting plasma.

Radio-interferometric (and sometimes optical) observations of some AGNs  
reveal jet features moving with apparent motions exceeding the speed of light. 
Such ``superluminal'' motion can be explained by a near-alignment of the jet 
with the line of sight combined with highly-relativistic motion of the radio plasma.
Photons emitted over a certain time interval reach the observer in a shortened 
time span as the plasma travels almost with the speed of light and thus
stays closely behind photons emitted towards the observer.
The apparent velocity can then exceed the speed of light.

If we denote the jet plasma velocity with $v\,=\beta_{\rm j} c\approx c$  ($c$ being the speed of light),
we can introduce the bulk Lorentz factor of the plasma with 
\begin{equation}
\Gamma_{\rm j}\,=(1-\beta_{\rm j}^{\,\,2})^{-1/2}.
\end{equation}
Emission from the jets is red or blue-shifted by the relativistic Doppler factor
\begin{equation}
\delta_{\rm j}\,=\,\frac{1}{\Gamma_{\rm j}(1-\beta_{\rm j} \cos{\theta})}
\label{df}
\end{equation}
with $\theta$ being the angle between the jet axis and the line of sight
as measured in the observer frame. 
A detailed study of a statistically complete, flux-density-limited sample of 
135 compact radio sources with the Very Long Baseline Array (VLBA) and 
the Mets\"ahovi Radio Observatory reveals relativistic motion with 
bulk Lorentz factors $\Gamma_{\rm j}$ between 1 and $\sim$40 \cite{Savo:10}.
\subsection{X-ray and Gamma-Ray Observations of kpc-Scale Jets}
\label{xg}
The Chandra X-ray observatory discovered X-ray emission from the kpc-scale jets 
and hotspots of a large number of radio-loud AGNs \cite{Harr:06}. The XJET 
web-site\footnote{http$://$hea-www.harvard.edu$/$XJET$/$}  lists 
the detection of X-ray emission associated with the jets from $\sim$120 AGNs.
The origin of the X-ray emission is still somewhat uncertain, especially for the higher-power FRII-type sources.

\begin{figure}[tb]
\begin{center}
\includegraphics[angle=0.,height=5cm]{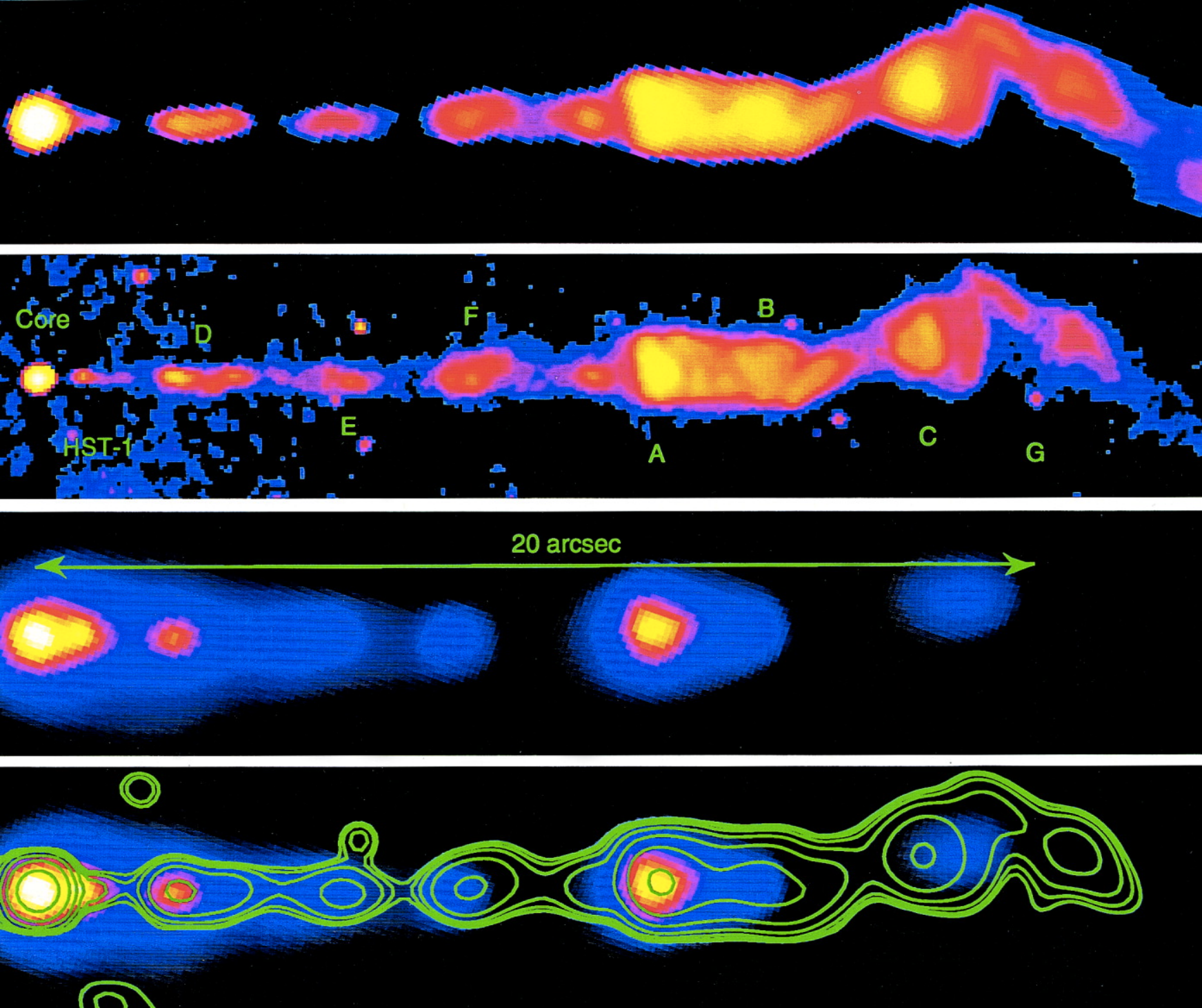}
\includegraphics[angle=0.,height=5cm]{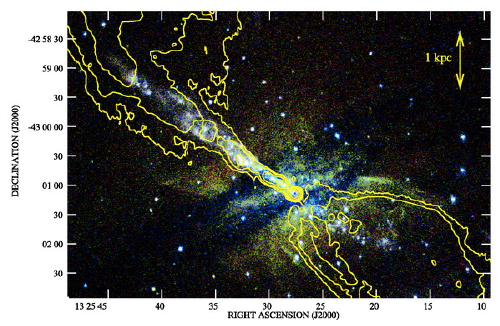}
\caption{Multiwavelength images of the radio galaxies M87 (left panel) and Cen~A (right panel).
 The left image shows from top to bottom the 14.435 GHz Very Large Array (VLA) image, 
 the Hubble Space Telescope Planetary Camera image in the F814W filter, and the 
 the Chandra X-ray (keV) image; the bottom panel shows the Chandra image overlaid with contours of the 
 Hubble Space Telescope image smoothed to match the Chandra point response function.  
 The Hubble Space Telescope and Very Large Array images 
 use a logarithmic color scale, and the X-ray image uses a linear scale.
The portion of the jet shown in the image has a projected length of $\approx$1.6 kpc.
The right image shows a false-color image of the radio galaxy 
Centaurus A as seen by Chandra (color image: 0.4-0.85 keV (red), 0.85-1.3 keV (green), and 1.3-2.5 keV (blue)) and
the VLA at 5 GHz (contours at  7$\times$ (1, 4, 16, ) mJy beam$^{-1}$).
The portion of the northern X-ray jet has a  projected length of $\approx$3~kpc.
The left panel is reproduced from Ref.~\cite{Mars:02},  \copyright 2002 American Astronomical Society,
and the right panel is reproduced from Ref.~\cite{Hard:07}, \copyright 2007 American Astronomical Society.
}
\label{Fknotty}
\end{center}
\end{figure}

For the low-power FRI-type sources, the X-ray emission form the kpc-jets can in most cases 
be explained as synchrotron emission from $\sim$TeV electrons gyrating in 
10-1000 $\mu$G magnetic fields. As electrons loose their energy on time scales of years, 
the X-ray bright spots imply in-situ particle acceleration. 
Well resolved jets show X-ray bright knots, sometimes spread over the extension of the jet  
(Fig.~\ref{Fknotty}).
The hypothesis of a synchrotron origin of the X-rays is supported by the rapid time variability of the X-ray flux from
the knots of the M87 jet \cite{Harr:06},
the radio to X-ray energy spectra which allow modeling with a single synchrotron component 
with a monotonic softening of the energy spectrum from the radio to the X-ray band (Fig.~\ref{FSEDs}, left panel), 
and the relative morphologies of the radio, optical and X-ray emission.  

\begin{figure}[tb]
\begin{center}
\hspace*{-0.5cm}
\includegraphics[angle=0.,height=7cm]{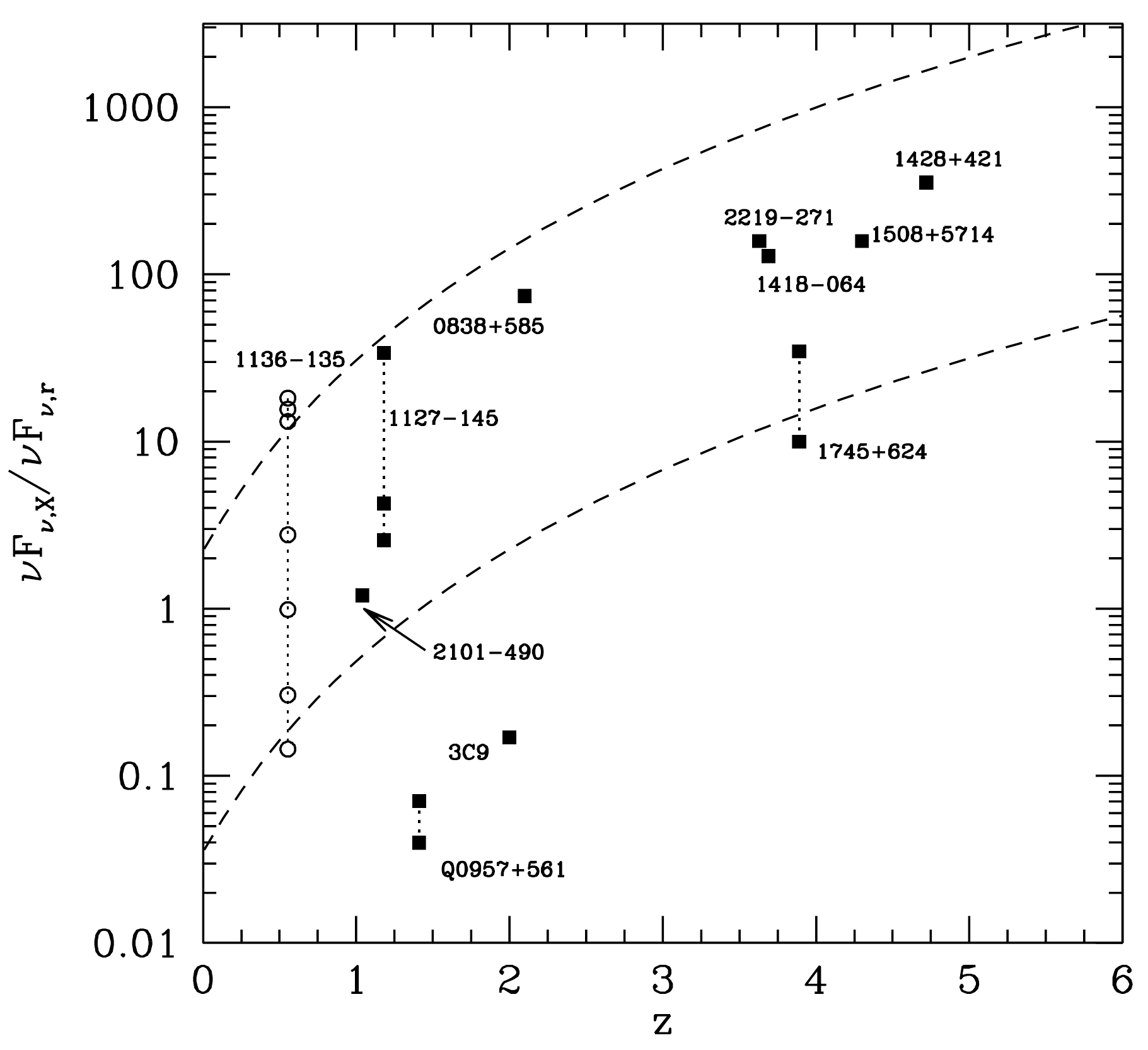}
\caption{Ratio of the X-ray and radio fluxes of high-redshift ($z>1$) jets discovered with Chandra following \cite{Cheu:04}.
For the nearby source  PKS 1136Ð135 at $z = 0.5$ the spread of the $z<1$ jet population is shown.
The data indicate that the ratio follows approximately the $(1+z)^4$-behavior predicted by the 
inverse-Compton/CMB model, albeit with a large spread (Courtesy of C.\ C.\ Cheung, 2012).}
\label{Fz}
\end{center}
\end{figure}

For the higher-power FRII-type sources, the radio to X-ray energy spectrum 
can often not be described with a single synchrotron component as the radio, optical and X-ray observations
imply a significant hardening of the energy spectrum between the optical and the X-ray band (Fig.~\ref{FSEDs}, right panel).
The presently favored explanation of the X-ray emission is that it originates
as inverse-Compton emission of mildly relativistic ($\sim$MeV) electrons 
embedded in a highly relativistic plasma moving with a bulk Lorentz factor $\Gamma\sim$10. 
In the co-moving reference frame of the jet plasma (primed variables), the 
mean frequency of cosmic microwave background (CMB) photons  is 
\begin{equation}
\nu'_{\rm CMB}=(1+z)\,\Gamma_{\rm j}\, \nu_0
\end{equation}
and the CMB energy density is
\begin{equation}
u'_{\rm CMB}\approx (1+z)^4\,\Gamma_{\rm j}^{\,\,2} \,u_{\rm CMB,0} 
\end{equation}
with the local present day values $\nu_0\approx 1.6 \times 10^{11}$~Hz  
and  $u_{\rm CMB,0}\approx 4\times10^{-13}$~erg~cm$^{-3}$.
The boosting of the mean energy and energy density proportional to  $\Gamma_{\rm j}$
and $\Gamma_{\rm j}^{\,\,2}$, respectively, explains the 
presence of a relatively strong inverse-Compton component in the X-ray band. 
The inverse-Compton/CMB model predicts an increase of the 
X-ray to radio brightness proportional to the energy density of the CMB which scales with $(1+z)^4$. 
The data indeed show such a trend for a small sample of sources (Fig.\ \ref{Fz}). 
However, the model faces several problems:
\begin{figure}[tb]
\begin{center}
\includegraphics[angle=0.,height=5.5cm]{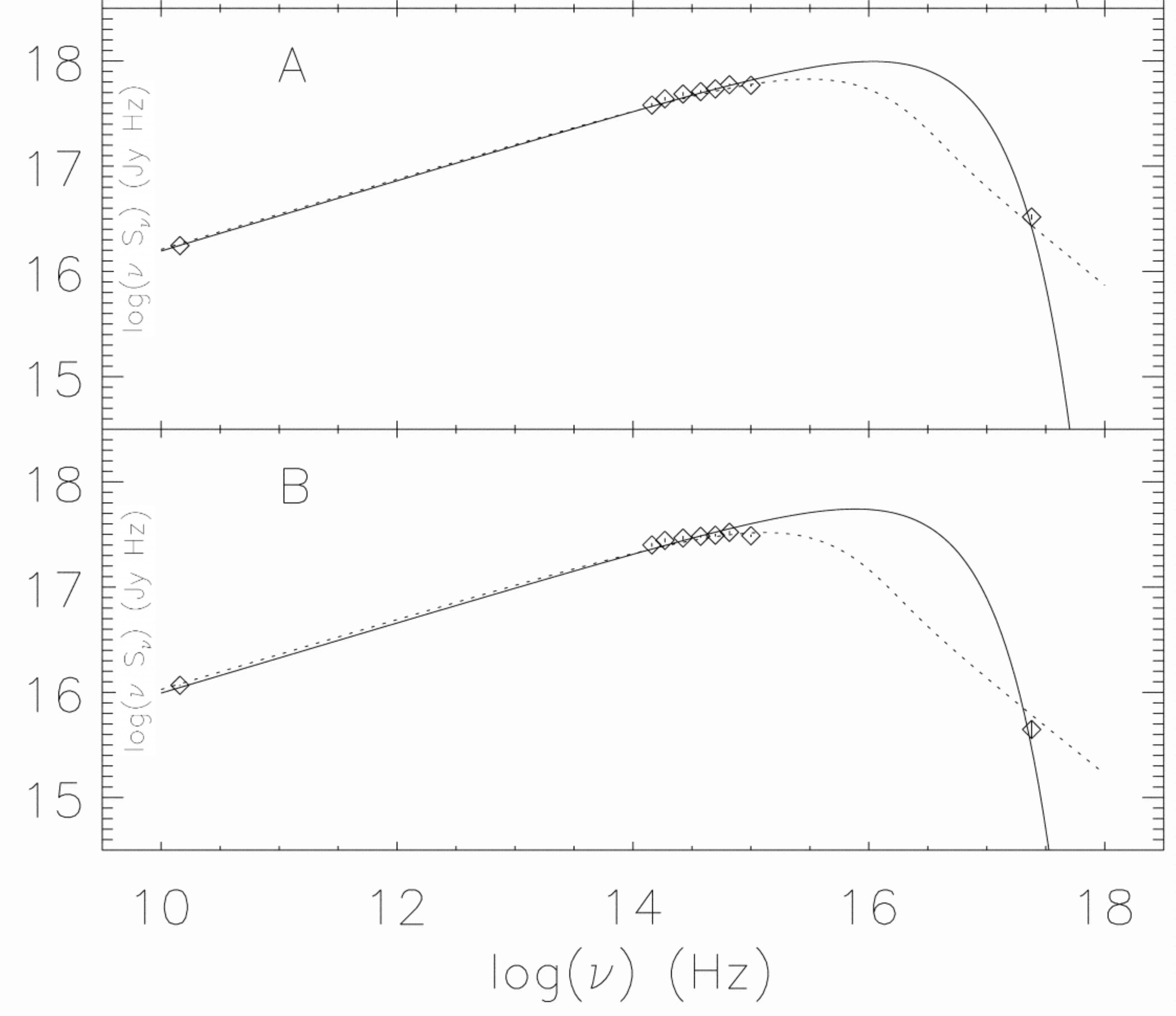}
\includegraphics[angle=0.,height=6cm]{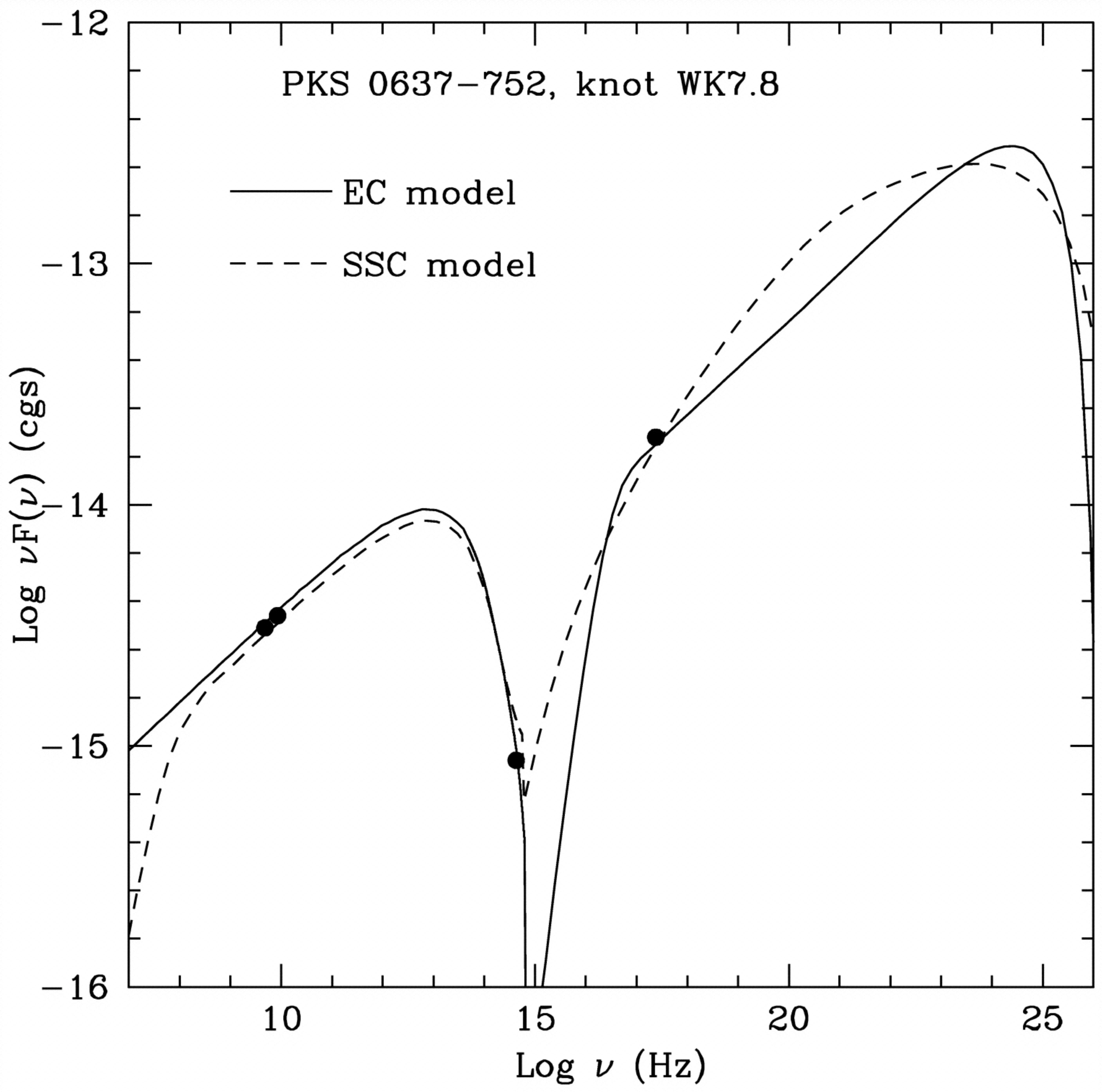}
\caption{Spectral energy distribution of two knots of the M87 jet (left panel, see the left lower-center panel of Fig.\ \ref{Fknotty} for the labels of the knots). The radio, optical and X-ray flux points can be explained
with a single population of non-thermal electrons (and/or positrons) with an energy spectrum that softens 
monotonically with energy. The dashed and dotted lines show synchrotron models.
The right panel presents the spectral energy distribution of the knot WK7.8 of the quasar PKS 0637-752 that shows evidence 
for two emission components, one component extending from radio to X-rays and another one extending 
from X-rays to gamma-rays. The authors favor the inverse Compton/CMB model (solid line) 
over the SSC model (dashed line) as the former requires less power than the latter (3$\times$10$^{48}$ ergs s$^{-1}$ compared to $>$10$^{49}$ ergs s$^{-1}$). 
The left panel is reproduced from Ref.~\cite{Mars:02},  \copyright 2002 American Astronomical Society,
and the right panel is reproduced from Ref.~\cite{Tave:00}, \copyright 2000 American Astronomical Society.
}
\label{FSEDs}
\end{center}
\end{figure}
for some of the sources the inverse-Compton/CMB model implies very small viewing angles and 
thus large physical source diameters on the order of hundreds of Mpc \cite{Harr:02};
furthermore, for some jets, the model parameters imply a rather high jet power 
($\sim$10$^{48}$~erg~s$^{-1}$) \cite{Atoy:04}.
With regards to jet models, the most interesting implication of the inverse-Compton/CMB model is the 
relativistic motion of the jet plasma at $\sim$kpc distances from the SMBH.
An alternative explanation for the observed X-ray emission is synchrotron emission from 
an additional high-energy electron population \cite{Harr:02}. 
However, in this case, the peak position of the second 
emission component at $\gg$keV-energies remains unexplained.

The {Fermi} LAT has recently detected emission from the core, and the northern and southern lobes 
of the radio galaxy Cen A \cite{Abdo:10c,Yang:12}. The emission can be explained as
inverse-Compton emission from electrons scattering photons of the CMB 
and maybe also higher-frequency photons. Whereas the synchrotron radio emission traces the combined 
properties of the radio-emitting electrons and the magnetic field strength, the inverse-Compton 
emission in the $\gamma$-ray band traces the high-energy electrons more directly. 
The combined data can also be used to set an upper limit on the magnetic field $B$ in the lobes of $B<$1$\mu$G.
It would be extremely interesting to obtain a$\gamma$-ray image with a substantially improved signal-to-noise ratio.
\subsection{Composition of AGN Jets}
\label{comp}
It is quite remarkable that we are still uncertain about the composition of AGN jets.
One of the reasons is the dominance of non-thermal continuum emission from jets leading to a 
lack of detected lines that give away the nature of the jet plasma.
Theoretical considerations suggest that the energy and momentum of the 
jets is initially (at distances of a few 10 $r_{\rm g}$)  dominated by electromagnetic 
energy (Poynting flux).  At larger distances (on the order of a few 100 or 
1000 $r_{\rm g}$) the electromagnetic energy is transferred to particles. 
The particles carrying the energy might be electrons and positrons or Interstellar Medium (ISM) 
processed through the accretion disk and/or entrained along the way. 
The jet may entrain additional ISM as it propagates, leading to a decrease 
of $\Gamma_{\rm j}$ with distance from the central engine.

For Flat Spectrum Radio Quasars (FSRQs), Sikora \& Madejski (2000)  argue that jets with $\Gamma_{\rm j}\sim10$ cannot be
dominated by cold pairs at their bases, as the inverse-Compton emission from the pairs scattering 
UV-radiation from the accretion disk would give rise to an unobserved 
soft X-ray emission component \cite{Siko:00} (see also \cite{Celo:93,Ghis:12}).

\subsection{AGN Feedback}
\label{feed}
\begin{figure}[tb]
\begin{center}
\includegraphics[angle=0.,height=6cm]{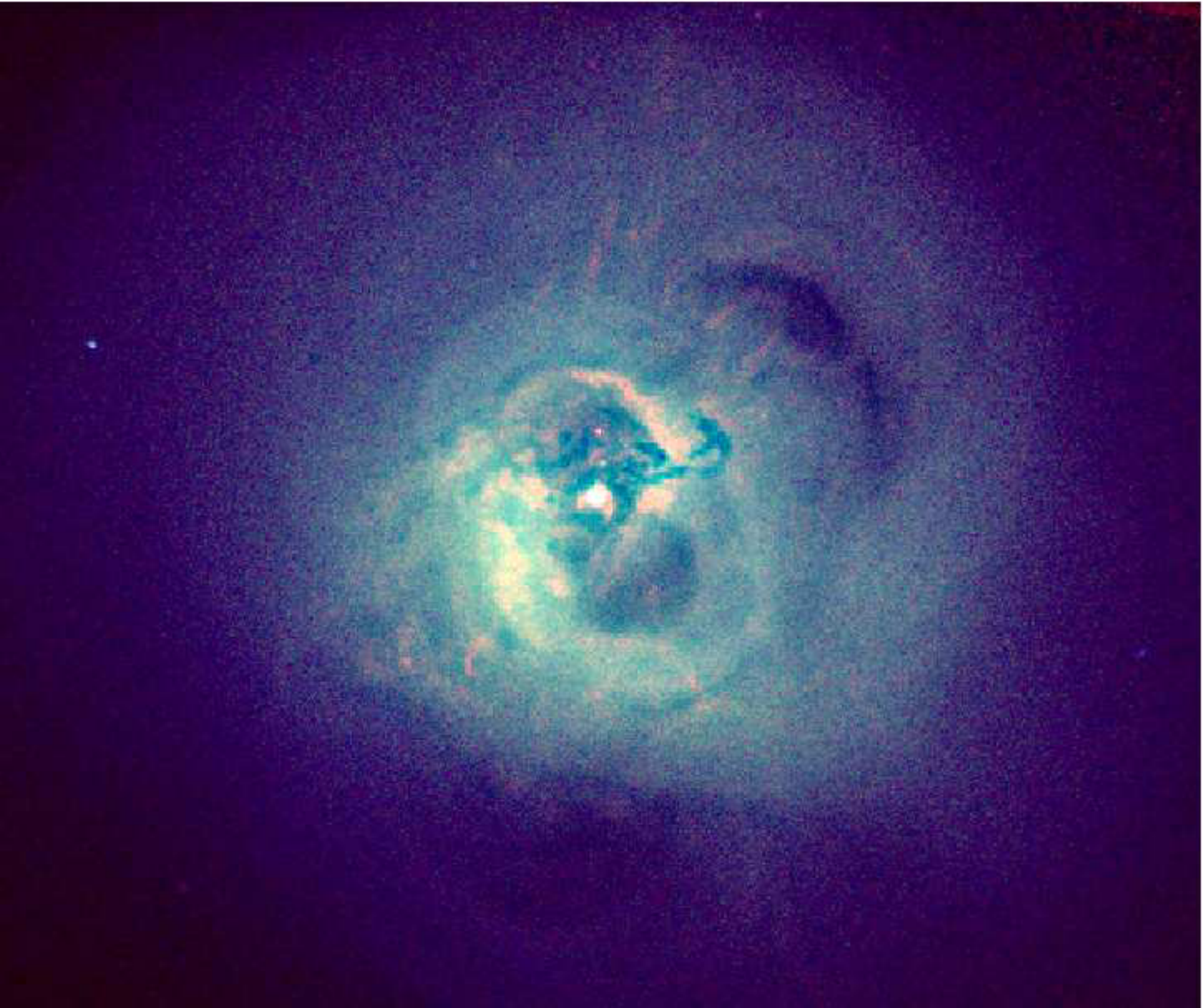}
\caption{A Chandra X-ray image of the Perseus galaxy cluster showing the emission from the
hot ICM. The X-ray brightness clearly shows two cavities carved out by the
radio plasma from the central radio galaxy NGC 1275. 
The ripples in the brightness distribution are explained as the result of sound 
waves propagating through the ICM. 
The opposite ends of the two radio bubbles are at a projected distance of $\approx$65 kpc, and the
bright ICM emission has a projected diameter of $\approx$360 kpc \cite[from][]{Fabi:06}.
Credits: NASA/CXC/IoA/A.Fabian et al..}
\label{Fperseus}
 
\end{center}
\end{figure}

\begin{figure}[tb]
\begin{center}
\includegraphics[angle=0.,height=7cm]{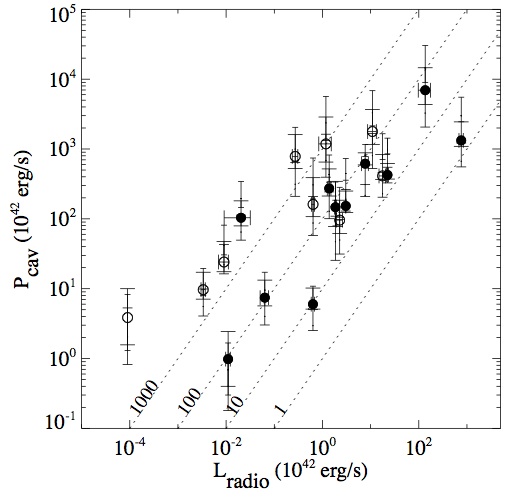}
\caption{The power required to blow cavities into the ICM as function of the power in 
the electrons that emit the observed radio emission. The former power exceeds 
the latter by an average factor on the order of 100 (diagonal lines), indicating that
the energy density of the radio plasma in the cavities is dominated by a ``dark 
component'' that has not yet been detected. The image is reproduced from 
Ref.~\cite{Birz:08},  \copyright 2008 American Astronomical Society.
}
\label{Fpower}
\end{center}
\end{figure}

Several X-ray observatories (first {\it ROSAT}, later Chandra and {\it XMM-Newton}) discovered 
large ``cavities'' in the hot X-ray bright gas of elliptical galaxies, galaxy groups, and galaxy clusters
associated with the radio lobes of AGNs \cite{Boeh:93,Cari:94} (see also Fig.\ \ref{Fperseus}).  The cavities are caused by AGNs 
inflating bubbles of radio plasma which displaces the hot X-ray bright interstellar or intracluster medium.
One can use these cavities to constrain the composition of the radio plasma in the cavities.
For galaxy clusters, the intracluster medium (ICM) pressure is well constrained by the spectroscopic 
X-ray images of the ICM. One can estimate the {\it p dV} work required to inflate the bubbles.
Combining this work with an estimate of the ages of the cavities (assuming they rise 
at the local sound speed), one can estimate the power required to inflate 
the bubbles of radio gas. Comparing the latter with the broadband radio power one can
estimate the ratio $k$ of the total jet power divided by the power carried by non-thermal 
electrons responsible for the observed radio emission. Such studies show $k\sim 100$ 
(Fig.\ \ref{Fpower})  and the presence of ``dark'' pressure-contributing components
(non-thermal low-energy electrons, non-thermal protons, or magnetic field).
Although AGN jets seem to have sufficient power to balance the radiative (Bremsstrahlung) 
cooling of the intracluster gas and to explain the lack of cold gas and freshly formed stars at the 
centers of galaxy clusters, the role of AGNs in heating the ICM is not well established \cite{McNa:12}.  
The jet power may not be transformed efficiently into ICM heat, and other processes (i.e.
efficient heat conduction owing to anisotropic transport properties) may dominate
the heating of cluster cores.

\subsection{Magnetohydrodynamic Simulations of Jets} 
The basic theoretical framework that is currently used to explain the observational appearance 
of radio-galaxies was introduced in the 1970s \cite{Rees:71,Long:73,Sche:74,Blan:74}. 
However, the predictive power of analytical jet models is limited owing to the non-linear processes 
taking place as jets propagate through the ambient medium. Recently, it has become possible to study jets with relativistic 3-D
hydrodynamic (HD) and magnetohydrodynamic (MHD) codes (see e.g.\ \cite{Mimi:12,McKi:12} and references therein).
The simulations can reproduce the observed morphologies of radio sources rather well.
Combined analytical and numerical studies show that certain ingredients can increase the 
stability of jets, i.e.\ magnetic fields, steep pressure gradients, a high density contrast 
between jet and external medium, fast motion, or sheath/shear velocity
outflow around the jet (e.g.\ \cite{McKi:09} and references therein).  
The codes use the one-fluid approximation and do not model the microscopic 
effects occurring in the astrophysical plasmas in detail. Particle in cell (PIC) simulations 
can shed light on some of the processes occurring when relativistic flows 
propagate through ambient media \cite[e.g.][]{Siro:11}.
\section{Emission from the Central Regions of AGNs}
\label{cores}
\subsection{Overview of Emission Components}
\label{core}
The radio, infrared, optical, UV, X-ray and $\gamma$-ray observations of AGNs can be explained with a 
single model of the central AGN region. In the paradigm, the different types of AGN result from different viewing 
angles towards the symmetry axis, and from the absence or presence of certain components \cite{Urry:95}. 
We briefly explain the most important emission components and their interpretation (see also Fig.~\ref{Fup}):
\begin{description}
\item[Emission from the Accretion Disk:] 
In some AGNs, emission from the accretion disk itself may have been detected. The feature, often referred to as the Big Blue Bump, can extend from the (AGN-frame) 
optical/UV band to the soft X-ray band and is believed to be thermal emission from the accretion disk 
\cite[e.g.][]{Czer:87,Kora:99}.

The accretion disk is partially covered by a corona of hot - yet still thermal - material. 
The hot corona Comptonizes some of the emission producing a high-energy tail 
extending into the hard X-ray regime. Alternative explanations of the origin of the 
hard X-ray emission include a hot inner flow \cite[e.g.][]{Ichi:77,Nara:02},  
the ``lamp-post model'' of an X-ray source illuminating the accretion disk 
from above \cite[e.g.][]{Henr:97,Malz:98}, and a structured multilayer corona \cite{Gale:79}.  

In the X-ray spectra of some AGN one detects a broad X-ray emission line at $\sim$6.4 keV in the AGN rest frame.
The line is believed to be fluorescence Fe K-$\alpha$ emission of iron atoms in the 
inner accretion disk excited by the hard X-rays from a source above the disk
\cite[e.g.][]{Fabi:89,Reyn:03,Ross:05,Ross:07}. Its shape results from the gravitational 
redshift incurred by photons climbing out of the black holes gravitational well, and the blue 
and redshift from the relativistic motion of the disk material. The analysis of the line shape can
be used to measure the black hole spin. For the Seyfert galaxy MCG6-30-15, the analysis of the 
Fe K-$\alpha$ line shape indicates a spin per unit mass of $a\,>$~0.987 in dimensionless units,
close to the theoretical maximum value \cite{Bren:06}. 

Radio interferometric observations at 230 GHz have recently achieved sufficiently good resolutions to 
resolve the accretion disk of Sgr A$^*$, the $4.5\times 10^6$ M$_{\cdot}$  SMBH at the center
of the Milky Way \cite{Doel:08}. The radio emission is polarized cyclo-synchrotron emission.
\begin{figure}[tb]
\begin{center}
\includegraphics[width=6cm,angle=-90]{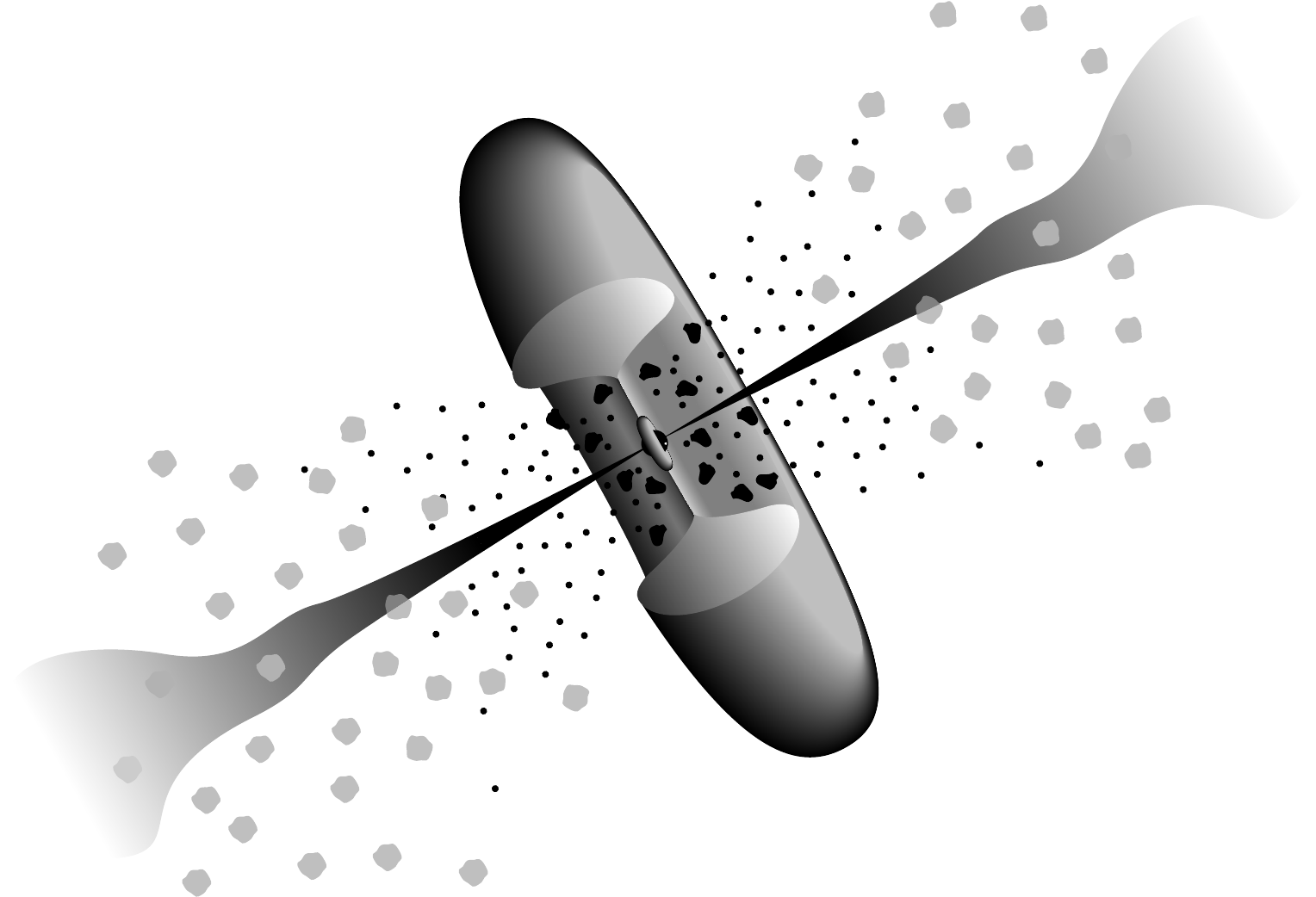}
\caption{\label{Fup} 
The figure shows a sketch of the central engine of an AGN and highlights several components 
that can explain most of the observed properties of AGNs (not to scale; example lengths in parentheses). 
The central black hole (for an $M\,=\,10^8\,M_{\odot}$ black hole, the Schwarzschild radius is 
$R_{\rm S} = 3 \times 10^{13}$ cm) is surrounded by an accretion 
disk ($\sim 1-30 \times 10^{14}$ cm). 
The broad emission lines originate in clouds orbiting above the disk (at $\sim 2-20 \times 10^{16}$ cm). 
A thick dusty torus (inner radius $\sim 10^{17}$ cm) or a warped accretion disk obscures 
the broad-line region (BLR) when the AGN is seen from the side; a hot corona above the 
accretion disk probably plays a role in producing hard X-rays; narrow lines are produced 
in clouds much farther from the central source ($10^{18} - 10^{20}$ cm). 
Radio jets (extending from $\sim 10^{15}$ cm to several times $10^{24}$ cm) 
emanate from the region near the black hole in the case of radio loud AGNs. 
The image does not show the broad and narrow absorption line producing winds. 
The launching region of the broad absorption line producing winds is uncertain. 
The narrow absorption line winds are believed to be launched at distances of $\sim$10$^{18}$ cm 
from the central engine at about the same distance from the central engine as the NLR clouds. 
The graph is reproduced from \cite{Urry:95}, \copyright 1995 Astronomical Society of the Pacific.
}
\end{center}
\end{figure}
\item[Continuum Emission from the Inner Jet:] Under favorable conditions, the accretion leads to the
formation of a highly relativistic collimated jet. The make-up of the jet is not well constrained, but
is believed to change from magnetic-field-dominated close to the central engine to particle
(electron and positron, or ions and electrons) dominated at $>$pc distances.
Shocks in the jet (e.g.\ from re-collimation of the jet or plasma instabilities) or reconnection
leads to the acceleration of electrons to GeV and TeV energies
(Lorentz factors up to a few times 10$^6$). The electrons emit low-energy synchrotron emission
and high-energy inverse-Compton emission. The latter comes from the electrons scattering synchrotron photons
(synchrotron self-Compton (SSC) emission) or external photons (external Inverse Compton (EIC) emission), such as 
BLR photons or photons from upstream or downstream regions of the jet.  
The continuum emission holds information about the innermost jet regions and can be used for
time-dependent studies of particle acceleration processes. 
\item[Emission Lines from Clouds:]
Some AGNs exhibit narrow and/or broad emission lines emitted by rather cold clouds 
of interstellar material orbiting the black holes at different distances \cite{Vero:00}. 
The broad line region (BLR) clouds have distances on the order of 10 light days 
from the SMBH; the narrow line region (NLR) clouds orbit the SMBH at distances of a few hundred parsecs.
The line centroids can be used to measure the redshifts and thus the distances of AGNs. 
Observations of the BLR emission and the continuum emission from the accretion 
disk can be used to estimate the mass of the central black hole based on reverberation mapping \cite{Pete:07,Kasp:07}. 
The technique combines the widths of the BLR lines (which constrain the orbital velocities of 
the emitting clouds) with measured time lags between variations of the continuum flux from the 
accretion disk and the BLR flux. 
As the BLR emission stems from reprocessing the continuum flux, 
the time lag can be used to estimate the distance of the BLR clouds from the central engine. 
The information about the velocity of the BLR clouds and the distance of the BLR clouds from 
the SMBH constrain the orbital parameters of the BLR clouds and thus the black hole mass.
The technique can be used for AGNs which are so far away that stellar orbits close to the black hole 
cannot be resolved. Two classical papers used the line luminosities as estimators of the accretion 
rate and studied the correlation of line luminosities with the kinematic power of the jet. 
Rawlings, S., \& Saunders \cite{Rawl:91} estimated the jet power based on the energetics of the radio lobes and correlated 
the estimated jet power with the [O III] NLR luminosities. 
The authors found a significant correlation between these two quantities for a sample of 
radio galaxies. Celotti et al.\ (1997) \cite{Celo:97} found a similar correlation between the  BLR luminosities 
and jet luminosities from very-long-baseline interferometry (VLBI) observations of radio loud AGNs \cite[see also:][]{Mara:03,Kawa:09}.
\item[Torus and Winds:]
Some of the differences between observational AGN classes can be explained by the presence 
of a dusty $\sim$1~pc diameter torus which can obscure accretion disk and BLR emission from view and emits 
reprocessed emission from the central engine in the infrared \cite{Elit:07,Nenk:08b}. 
At the inner edge of the torus, the AGN continuum emission destroys the dust, ionizes the atoms and 
creates the material making up the BLR and X-ray obscuring clouds.
The presence of blue-shifted broad (BAL) and narrow absorption lines (NAL) in the optical, 
UV, and X-ray regime show evidence for fast AGN outflows or winds \cite{Chel:07}. 
While the NAL outflows are believed to be launched at $\sim$pc distances from the central 
engine and are largely radiatively driven, the location and the driving mechanism of the 
fast (0.2 $c$) BAL outflows are still highly uncertain. 
\end{description}
At larger viewing angles, the gas torus  conceals the BLR and only narrow lines are observed, resulting in
Type-2 Seyferts, and narrow line FRI and FRII galaxies. Closer to the line of sight, the torus does not
obscure the BLR anymore and the AGNs appear as Type-1 Seyferts, radio quiet quasi-stellar objects (QSOs),
and broad line steep spectrum radio quasars (SSRQs) and FSRQs.
For  viewing angles $< 10^{\circ}$, the relativistically beamed non-thermal continuum 
emission from the relativistic jet dominates and the object is a blazar.
BL Lac objects are a sub-class of blazars either without detected emission lines or with lines with
a rest-frame equivalent width smaller than 5 \AA ~\cite{Stic:91,Stoc:91}. Recently, this classification 
scheme has been criticized on the grounds that it introduces strong selection effects
\cite{Ghis:11,Giom:12}. 
\subsection{Recent Blazar Observations with the {Fermi} Space Telescope and Imaging Atmospheric Cherenkov Telescopes}
\begin{figure}
\begin{center}
\includegraphics[angle=0.,width=7.6cm]{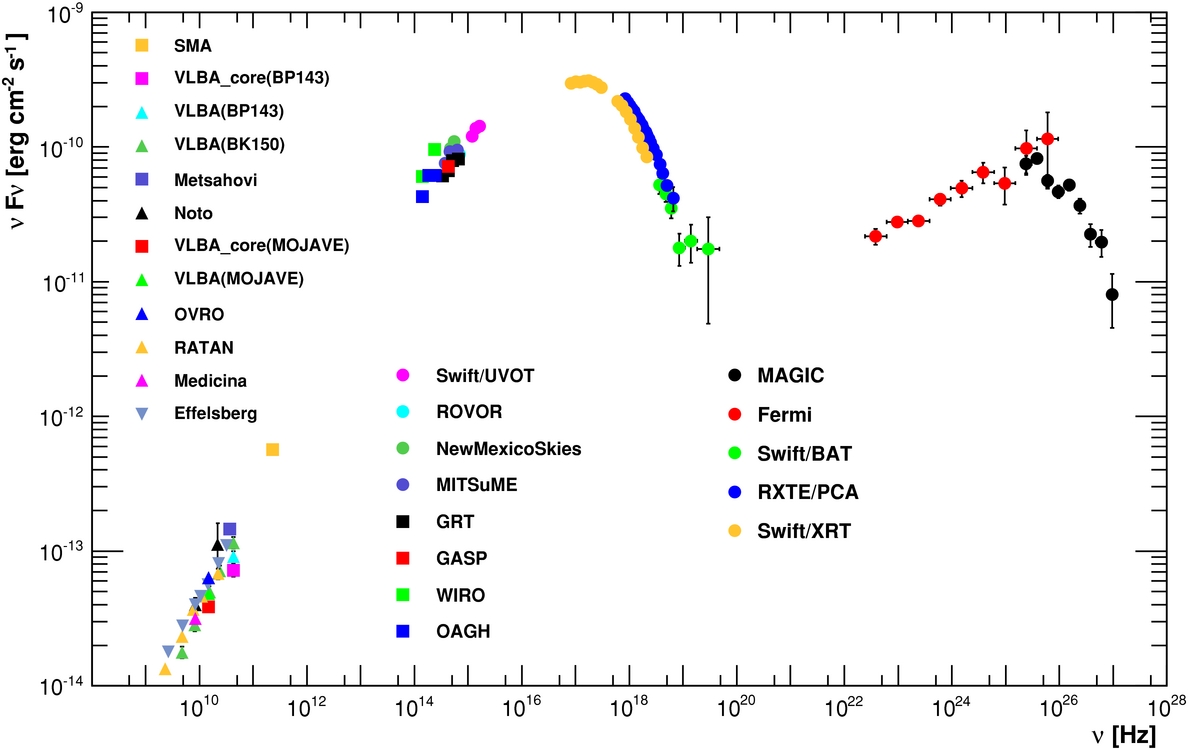}
\includegraphics[angle=0.,width=7.3cm]{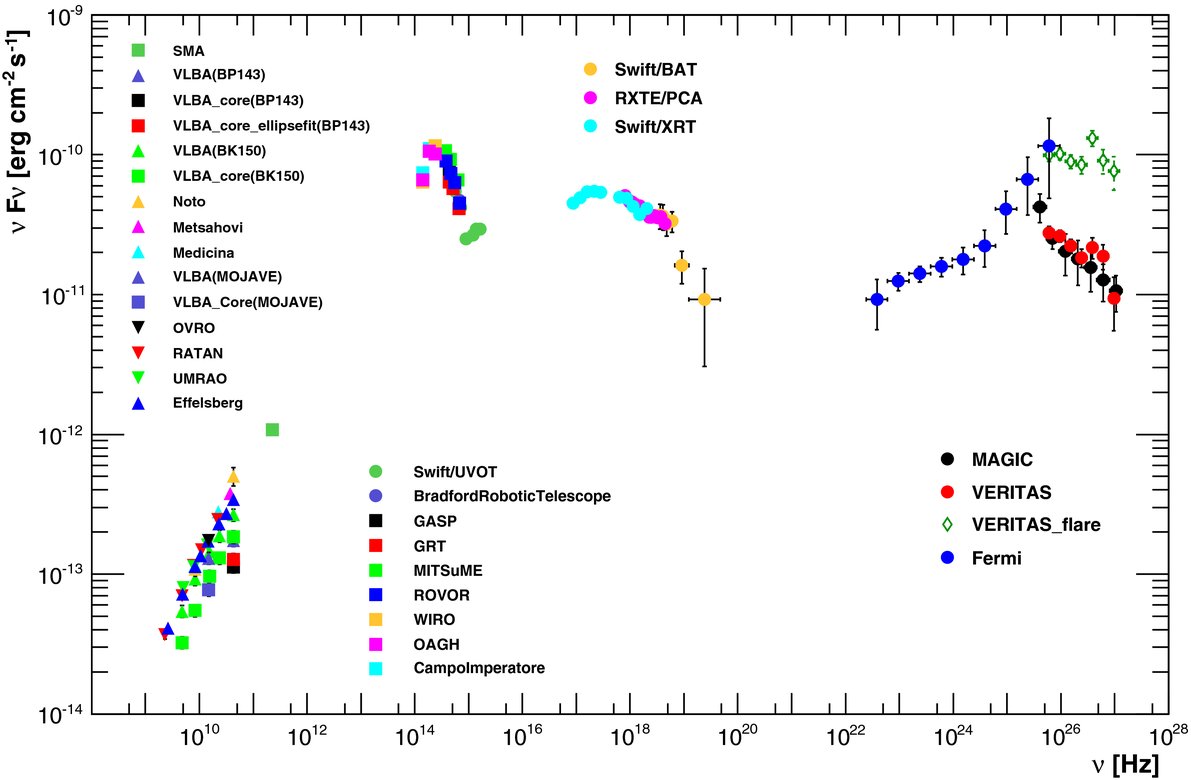}
\caption{Broadband spectral energy distribution of the HSP blazars Mrk 421 (left panel, data taken 
between January 19, 2009 and June 1, 2009) and Mrk 501 (right panel, data taken 
between March 15, 2009 and August 1, 2009).
The images are reproduced from Refs.~\cite{Abdo:11a} (left panel) and \cite{Abdo:11b} (right panel), 
 \copyright 2011 American Astronomical Society.
}
\label{David}
\end{center}
\end{figure}

In recent years $\gamma$-ray astronomy has made spectacular progress. The space-borne {Fermi} Large Area 
Telescope (LAT) \cite{Atwo:09} and ground-based Imaging Atmospheric Cherenkov Telescopes (IACTs)  
\cite{Ahar:08} achieve sensitivities of about $10^{-12}$ erg cm$^{-2}$ s$^{-1}$.
The Fermi LAT covers the 100 MeV-300 GeV energy range. With its large field of view ($\sim$2.4 sr) it is 
well suited to monitor the entire sky on a regular basis.  IACTs cover the 50 GeV-30 TeV energy range. 
With large detection areas on the order of $10^5$ m$^2$ they are well suited to measure 
short-term ($<$1 min) flux and spectral variability. Between Fermi and the ground-based 
Cherenkov telescopes, it has become possible to sample the complete 
inverse Compton emission component of many blazars (Fig.\ \ref{David}).
  
The $\gamma$-ray telescopes detect mostly blazars and make it possible to study the properties of AGN jets
on short time scales. In the following we use the abbreviations LSP, ISP and HSP to denote 
low synchrotron peaked ($\nu_{\rm peak}^{\rm S}<10^{14}$~Hz), 
intermediate synchrotron peaked ($10^{14}$~Hz$<\nu_{\rm peak}^{\rm S}<10^{15}$~Hz), and
high synchrotron peaked ($\nu_{\rm peak}^{\rm S}>10^{15}$~Hz) blazars.
In the earlier literature, the reader will often find the abbreviations LBL, IBL, and HBL.
These names denote the BL Lac subclasses of LSP, ISP, and HSP blazars, respectively.  

The large sample of blazars detected by the {Fermi}-LAT can be used to study the cosmic evolution of 
blazars as well as the intrinsic properties of blazars. The cosmic evolution of blazars depends on the 
cosmic evolution of SMBHs, the accretion history of these SMBHs, and the history of the
radiative efficiency of the accreting SMBHs. 
The recently published second {Fermi} LAT AGN catalog \cite{Nola:12} lists 1017 
sources with an AGN association,  and a ``clean sample'' of 886 sources with unambiguous AGN counterparts.
The latter sample includes 395 BL Lacs, 310 FSRQs, 157 candidate blazars of unknown type,  
eight misaligned AGNs, four narrow-line Seyfert 1 galaxies, ten AGNs of other types,  and two starburst galaxies.
The redshift distribution ($dN/dz$) of detected BL Lacs peaks at $z\approx0.2$ and extends to $z=1.5$; the
distribution of FSRQs peaks at $z\approx 1.1$ and extends to $z=3.2$.

\begin{figure}
\begin{center}
\includegraphics[angle=0.,height=5cm]{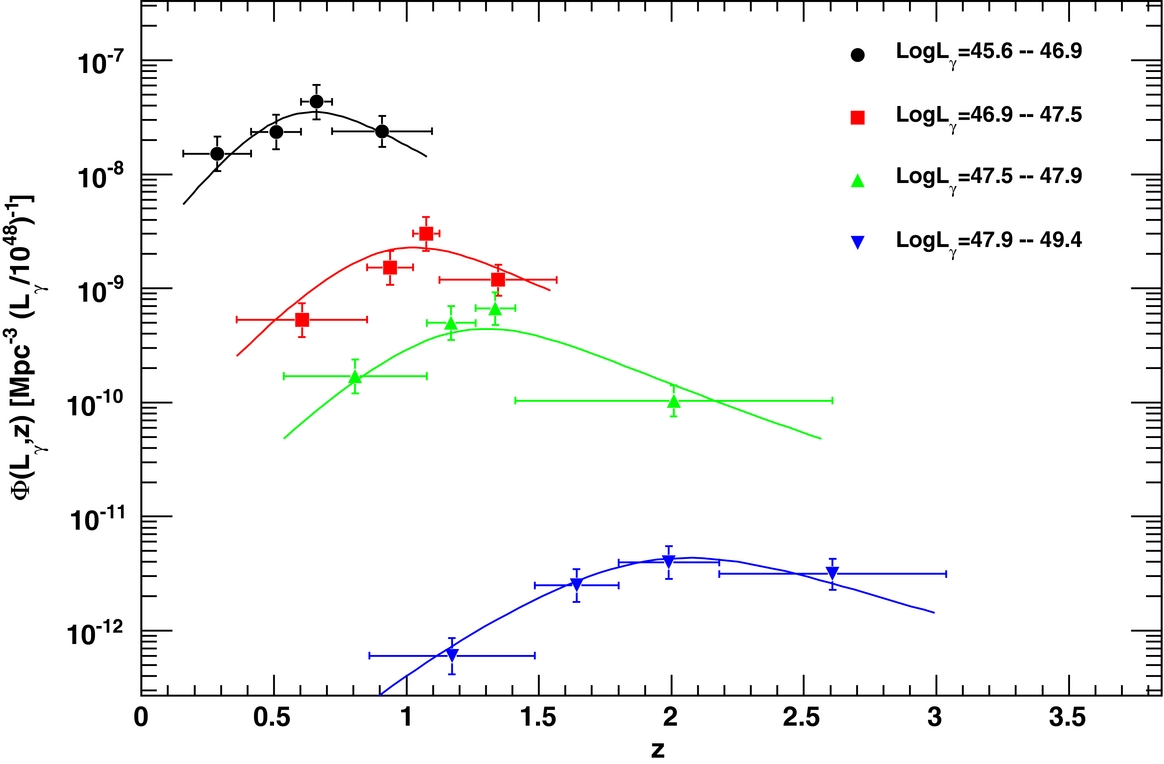}
\includegraphics[angle=0.,height=5cm]{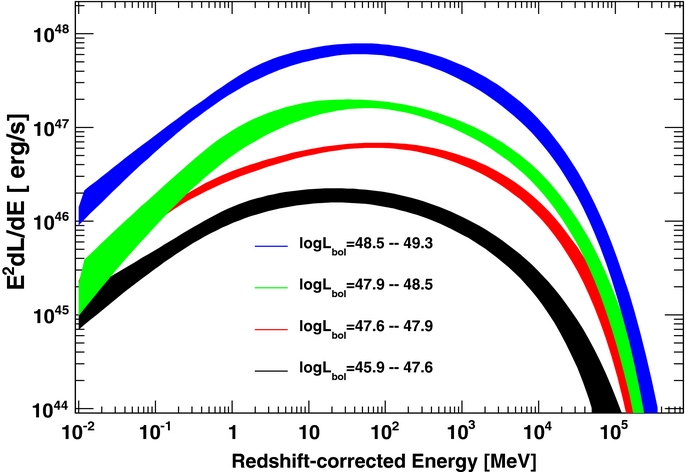}
\caption{Left panel: space density of Fermi-LAT detected FSRQs for different luminosity classes indicated an 
``inverted hierarchy'' with higher (lower) luminosity objects peaking at higher (lower) redshifts.  
 Right panel: average rest-frame spectral energy distributions for four FSRQ luminosity classes. 
 The bands show the $1\sigma$ 
 confidence level regions \cite[from][]{Ajel:12}. The data do not indicate a correlation of the energy at which the
 spectral energy distribution peaks and the $\gamma$-ray luminosity.
 The images are reproduced from Ref.~\cite{Ajel:12}, 
 \copyright 2012 American Astronomical Society.
 }
\label{Fajel}
\end{center}
\end{figure}

Ajello et al. (2012) use a complete sample of FSRQs detected during {Fermi}'s first year of
operation to study the luminosity function (LF) and its cosmic evolution \cite{Ajel:12}. 
The FSRQ number density grows up to luminosity dependent redshifts of between
0.5 and 2.0 and declines thereafter. The sources show an ``inverted evolution'' with
lower power sources peaking at lower redshifts (later times) than their 
higher-power counterparts (Fig.\ \ref{Fajel}, left panel). 
The authors study the rest-frame  spectral energy distribution 
(SED, energy flux per logarithmic energy interval, $E^2 dN/dE$) of the sources and find that
the peak of average SEDs determined for different luminosity bins is independent of the 
AGN luminosity (Fig.\ \ref{Fajel}, right panel). The earlier detection of a 
luminosity-hardness correlation (sometimes referred to as the ``blazar sequence'')  
\cite{Foss:98,Ghis:98} may have resulted from selection effects. 
Combining the inferred SED shapes with the LF, the authors predict that FSRQs contribute 
$\sim$10\% to the isotropic {Fermi} $\gamma$-ray background.
Assuming that the jet Bulk Lorentz factors $\Gamma_{\rm j}$ follow a power law distribution
$dN/d\Gamma_{\rm j}\propto \Gamma_{\rm j}^{\,\,k}$ 
over the range from 5 to 40 and that the $\gamma$-rays originate as SSC emission, the authors can derive
the distribution of intrinsic jet luminosities ${\cal L}$ from the distribution of apparent luminosities $L$ 
(for SSC emission, $L\propto \delta_{\rm j}^{\,\,p} {\cal L}$ with $p\,=\,4$).
The analysis implies a power law index $k$ of -2.0$\pm$0.7, 
a mean Lorentz factor of the {Fermi}-detected FSRQs of $\approx$12,  
that most sources are seen from within $5^{\circ}$ of the jet axis, 
and that the detected $\gamma$-ray loud FSRQs represent 0.1\% of the unbeamed parent population.
The distribution of $\Gamma_{\rm j}$ is in good agreement with those inferred from radio observations
on pc-scales \cite{Savo:10}. The finding may imply that both, the radio and the $\gamma$-ray emission comes 
from $\sim$pc-distances from the central engines, or, that the jet Lorentz factors are the same at the 
scales probed by the radio and $\gamma$-ray observations.

Ghisellini et al.\ (2012)  study the correlation between the BLR luminosity $L_{\rm BLR}$ and the 
$\gamma$-ray luminosity $L_{\gamma}$, both measured in units of the Eddington accretion rate,
and find a correlation albeit with a large scatter \cite{Ghis:12}.
The result reinforces the earlier finding that accretion rate 
(and thus $L_{\rm BLR}$ and jet power (proportional to $L_{\gamma}$) are correlated, but shows 
that it is even valid when both are normalized to the mass of the black hole. 

Giommi et al.\ (2012) report on a Monte Carlo study that shows that selection effects can strongly affect the outcomes
of BL Lac and FSRQ studies \cite{Giom:12}. 
Their work indicates that selection effects alone can 
produce an apparent correlation between the peak energy of the synchrotron SED and the luminosity of the sources.
The analysis suggests that powerful BL Lac-type objects may exist, but cannot be identified as such owing to 
the relative weakness of the lines. In their model, all sources are qualitatively the same, but are identified 
as different sources owing to the relative strength of several basic emission components, i.e.\ 
the Doppler boosted radiation from the jet, the emission from the accretion disk, the BLR, 
and the light from the host galaxy. The authors suggest that -- using the standard definitions -- 
BL Lacs and FSRQs may not be the beamed versions of FRI and FRII, respectively.
Ghisellini et al.\ (2011) offer a definition of BL Lacs and FSRQs based on the broad line region
luminosity in units of the Eddington luminosity ($L_{\rm BLR}/L_{\rm Edd}$
below and above $5\times 10^{-4}$, respectively) \cite{Ghis:11}, which might result in a cleaner identification 
of the beamed versions of FRI and FRII galaxies.

IACTs and the {Fermi} LAT have been used for extensive multiwavelength campaigns.
For HSP blazars, the broadband SEDs \cite[e.g.][]{Kraw:01,Kono:03} and fast flux variability
\cite{Gaid:96,Ahar:07,Albe:07,Bege:08} imply very high relativistic Doppler factors $\delta_{\rm j}$ 
(Equation (\ref{df}))
on the order of 50. The broadband SEDs can largely be fit with one-zone SSC models 
\cite[e.g.][]{Acci:11,Alek:12} with notable exceptions \cite[e.g.][]{Kraw:04,Hess:12}. 
Snapshot and time-dependent modeling indicates particle dominated emission zones with 
electron and positron energy densities $u_{\rm e^{+/-}}$ exceeding the magnetic field energy 
density $u_{\rm B}\,=\,B^2/8\pi$ by typical factors of $\sim 100$ \cite{Kraw:01,Acci:11}. 
Models with additional target photon fields (and with additional degrees of freedom) 
make it possible to fit the data closer to equipartition ($u_{\rm e^{+/-}}\sim u_{\rm B})$ \cite{Geor:04,Ghis:05}.

LSP and ISP blazars seem to require EIC models. Abdo et al.\ (2010) discuss the broadband SEDs 
of 48 $\gamma$-ray bright {Fermi} blazars \cite{Abdo:10}. The authors find that one-zone SSC models cannot
fit the SEDs of most LSP and ISP (low frequency peaked BL Lac objects and FSRQs) sources
\cite[see also:][]{Giom:12x}.

\begin{figure}[tb]
\hspace*{0.7cm}
  \includegraphics[height=.5\textheight]{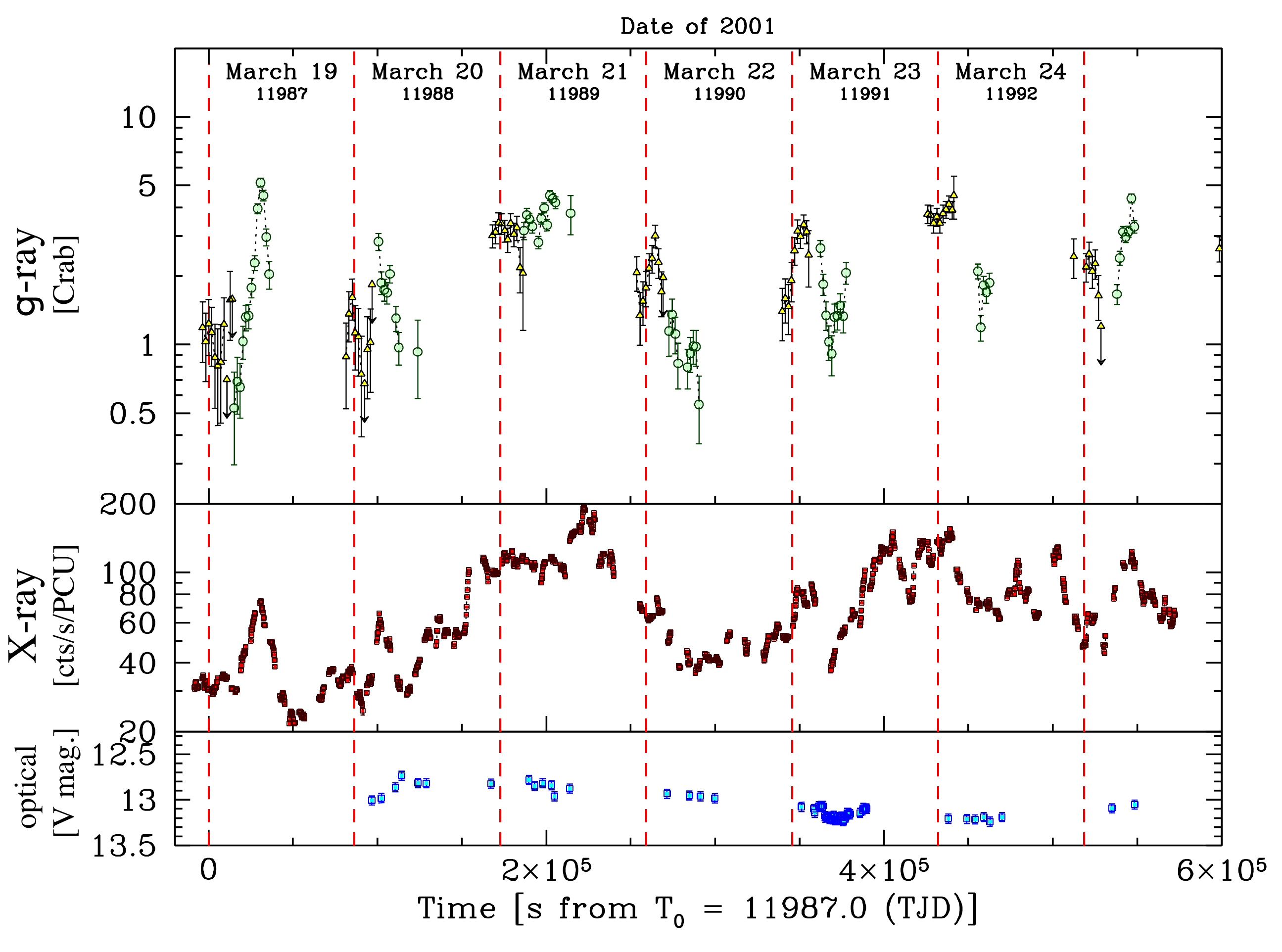}
  \caption{Results from 2001 Rossi X-ray Timing Explorer ({\it RXTE}) 2-4 keV X-ray and
	Whipple (full symbols) and HEGRA (open symbols) $\gamma$-ray observations 
	of Mrk 421 in the year 2001 \cite[from][]{Foss:08}. 
	IACTs achieve excellent sensitivities on short time scales owing
	to their large collection areas. Careful study of the light curves shows that the 
	X-ray and TeV $\gamma$-ray fluxes are correlated for some flares but not for all.
	Reproduction of the figure with kind permission of the authors. \label{giovanni}}
\end{figure}

%
%
Broadband observations promise to give insights into the mechanisms responsible for the observed flares.
There is good evidence for a correlation of the radio and $>100$ MeV $\gamma$-ray luminosities of 
statistical samples of blazars \cite{Ghir:11,Acke:11,Niep:11,Giom:12x}, and to some degree of 
the X-ray and VHE  (Very High Energy, $>$100 GeV) $\gamma$-ray fluxes of HSP blazars 
(e.g.\ Fig.\ \ref{giovanni} and \cite{Kraw:00,Foss:08}). 
Marscher et al.\ (2008) \cite{Mars:08} and Abdo et al.\ 2010 \cite{Abdo:10b} report the detections of correlated 
polarization swings and $\gamma$-ray flares.
Unfortunately, the interpretation of some of the data is rather ambiguous as the sources exhibit 
a wide range of different behaviors; as a consequence, the statistical significance of the observed features is often 
rather limited. Unambiguous conclusions would require the observation of a statistical sample of 
sources with dense sampling in the temporal and waveband domains over many years.

The continuum emission from blazars is most commonly explained in the framework of electron
and possibly positron acceleration in internal or external collision-less shocks. 
The accelerated leptons radiate synchrotron radio to X-ray and inverse-Compton X-ray to gamma-ray emission. 
Unfortunately, attempts to verify that Fermi-type acceleration energizes the leptons based on observations of 
predominantly clockwise or counterclockwise loops in the flux vs.\ spectral 
index plane \cite[e.g.][]{Kirk:99} failed due to the rather chaotic behavior 
of the observed sources \cite[e.g.][]{Taka:00,Gars:10}. 

Some authors consider hadronic models in an attempt to explain the $\gamma$-ray emission from
blazars, but such models require the acceleration of protons to $>10^{18}$~eV energies 
and magnetic fields typically on the order of $\sim$50~G with very high energy densities 
to accelerate and confine the protons \cite{Mann:93,Muec:03,Ahar:00,Reim:12,Boet:12}.
\subsection{Models of the Accretion Flow and Jet Formation}
\label{flows}
A satisfactory model of the central engines of AGNs should explain 
(i) how matter and magnetic fields are transported towards the accretion disks of the SMBHs;
(ii) which types of accretion disks occur in nature and how they work; 
(iii) which physical mechanisms are responsible for accretion disk state transitions and flares;  
(iv) how the individual emission components are produced,  (v) how jets form, transform, accelerate 
and/or decelerate at different distances from the SMBHs; (vi) how AGNs interact with their 
environment.
The question of how matter and magnetic fields move towards the central engine (the feeding 
problem) will not have a simple answer as it will depend on the cosmic epoch and the type 
and evolutionary state of the host (a single galaxy or a galaxy inside a galaxy cluster). 
The main challenge is to explain how the matter can shed all but a tiny fraction of its initial 
angular momentum while it goes through different phases and moves from
$\sim$kpc distances to the accretion disk \cite[e.g.][]{Hopk:12,McKi:12}. 
One of the open questions 
concerns the feeding of magnetic fields with a preferred polarity into the accretion flow, 
as such magnetic fields can suppress plasma instabilities in the disk, and can explain 
the presence of a strong single polarity magnetic field in the surrounding of the black hole, i.e.\ in the
plunging region between the event horizon and the innermost circular stable orbit (ISCO). 
Such magnetic fields are required in some models of accretion and jet formation.

Accretion disks transform gravitational energy of matter into electromagnetic and mechanical energy.
Shakura \& Sunyaev (1973) introduced a model for a geometrically 
thin (with a thickness $H$ at radius $r$ such that $H(r)/r\ll 1$), 
optically thick accretion disk \cite{Shak:73}.  Assuming that the disk matter orbits the black hole on
circular geodesics, that there is no torque at the ISCO, that the disk radiates away 
all the dissipated energy, and that no heat is transported in radial direction, the radial structure of the 
disk is entirely determined by mass, energy, and angular momentum conservation \cite{Shak:73,Page:74}. 
Using a prescription for the viscosity of the disk, the horizontal 
disk structure can be inferred. Although it was suspected that magnetic turbulence 
caused by the differential rotation of the accretion disk material was responsible 
for the viscosity, it was only in 1991 that Balbus \& Hawley  identified 
the magneto-rotational instability (MRI) as the driving instability based on 
numerical simulations \cite{Balb:91}.

A number of authors discuss alternative accretion flows. 
Ichimaru (1977) describes a two-state model to explain 
two qualitatively different emission states of the X-ray binary Cygnus X-1 \cite{Ichi:77}. 
Whereas the high-soft state corresponds 
to the geometrically thin, optically thick radiatively efficient accretion disk of Shakura \& Sunyaev, the low-hard state 
corresponds to a geometrically thick, optically thin radiatively inefficient accretion flow (RIAF). 
In the latter case, a thermal instability of the disk plasma develops when dissipative heating exceeds the 
radiative cooling causing the disk to puff up. Narayan et al.\ (1994) discusses 
a self-similar geometrically thick RIAF flow, the advection dominated accretion flows (ADAFs), 
in which the gas orbits the black hole with a velocity well below that of Kepplerian orbits \cite{Nara:94}. 
The authors remark that such flows can form for low accretion rates when the 
flow is optically thin, or for very high accretion rates when 
the flow is optically thick and the cooling time of the plasma is much longer than the 
accretion rate. Variations of ADAFs include convection-dominated accretion flows (CDAFs) 
\cite{Nara:00,Quat:00}, and advection-dominated inflow-outflow solutions (ADIOSs) \cite{Blan:99,Bege:12}.

The jet is probably launched by the combined effect of thermal pressure, centrifugal forces, 
and the Blandford-Znajek  process. The latter involves the conversion of the rotational energy of 
a black hole spinning in the magnetic field anchored in the accretion disk 
into electromagnetic energy  \cite{Blan:77}.  In the presence of a favorably shaped outflow channel 
(formed by a geometrically thick accretion disk or by a less collimated wind),
the flow can accelerate owing to magnetic pressure gradients. Energy conservation dictates 
that the terminal Lorentz factor of the jet obeys $\Gamma_{\rm j}^{\infty}< \sigma_0$ 
with $\sigma_0$ being the magnetization (ratio of electromagnetic to particle energy densities) 
at the base of the jet, so that $\sigma\gg 1$ is required to explain $\Gamma_{\rm j}^{\infty}\gg 1$. 
 
We would like to know which accretion flows occur in nature, which flow properties lead  
to the observed phenomenology, and how the observed jets form. Attempts in this direction include the identification of the
radio quietness and loudness (the absence or presence of a jet) with geometrically thin 
and geometrically thick accretion flows, respectively. 
Some authors explain the difference between BL Lacs and 
FSRQs by invoking radiatively inefficient (with weak BLR emission) accretion flows for the former and 
radiatively efficient (strong BLR emission) accretion flows for the latter sources.  Unfortunately, none of these 
associations is firm at the time of writing.

Recently it has become possible to employ 2D and 3D general relativistic magnetohydrodynamic (GRMHD) simulations 
with sufficient resolution to test some of the assumptions underlying the analytical and semi-analytical models.
Most simulations neglect radiative transfer of heat owing to computational limitations. 
Such simulations have been used, for example, to test the assumption of zero torque at the ISCO (and zero energy 
dissipation of the disk plasma within the ISCO). The results indicate that the zero-torque approximation introduces 
rather small errors, i.e.\ it underestimates the emitted luminosity by $\sim$5\% \cite{Nobl:11,Penn:12}. 
McKinney et al.\ (2012) studied rather thick ($H/r\sim 0.3$) accretion flows with large-scale dipole 
and quadrupole magnetic fields and obtained two
interesting results \cite{McKi:12}: 
(i) the structure of the accreted magnetic field is decisive for the formation of a collimated relativistic
outflow; an accretion disk with a dipole magnetic field geometry does produce a jet, but disks fed by plasma
without an ordered magnetic field or with higher moment magnetic field geometries do not 
\cite[see also:][]{McKi:07,Beck:08}; (ii) somewhat unexpectedly, the jets are stable even 
though the toroidal component of the magnetic field that accelerates the jets could disrupt the flow owing 
to helical kink and screw modes. Several effects -- including  gradual shear, stabilizing  sheaths, or 
sideways expansion -- may be responsible for stabilizing the outflow.

As ordered magnetic fields are needed for the production of jets, McKinney et al.\ (2012) employ 
3D GRMHD simulations to study a geometrically thick flow supplied with strongly magnetized plasma \cite{McKi:12}.
They find that for rapidly spinning black holes toroidal magnetic fields can lead to large patches of
single-polarity poloidal magnetic fields threading the black hole enabling the transformation of 
rotational energy of the black hole into Poynting flux energy. 
Strong poloidal magnetic fields build 
up in the inner region of the disk and compress it  into a geometrically thin accretion flow in which 
the strong poloidal magnetic field suppresses the MRI. As mentioned above, further studies are needed to
understand which accretion flows are actually realized in nature.

\section{The Cosmic History of Black Hole Accretion}
\label{ev}

Most current black hole formation models tell us that the first black hole seeds emerged at $z$$\gsim$15. While the exact
mechanism is not known, there are several prevailing theories (see the comprehensive reviews by 
M. Rees \cite{rees78} and M. Volonteri \cite{volonteri10a} for more details). 
One possibility is that the first black holes were the remnants of the first generation of stars (Population III stars)
that resulted from the gravitational collapse of primordial ultra-low metallicity gas. 
These black holes formed at $z$$\sim$20 and have typical masses $\sim$100-1,000~$M_\odot$.
This scenario has problems explaining the very high masses, 
of $\sim$10$^9$M$_\odot$, estimated for supermassive black holes in $z$$>$6 optically-selected quasars \cite{willott10b,mortlock11}. 
Alternatively, the first black holes could have formed directly as the result of gas-dynamical processes. It 
is possible for metal-free gas clouds with $T_{\textnormal vir}$$\gsim$10$^4$K and suppressed $H_2$ formation to collapse 
very efficiently \cite{bromm03}, possibly forming massive black hole seeds with M$\sim$10$^4$-10$^5$M$_\odot$ as early 
as $z$$\sim$10-15. If instead the UV background is not enough to suppress the formation of $H_2$, the gas will fragment 
and form ``normal'' stars in a very compact star cluster. In that case, star collisions can lead to the formation of a very 
massive star, that will then collapse and form a massive black hole seed with mass $\sim$10$^2$-10$^4$M$_\odot$
\cite{devecchi09}.

Given the current typical masses of 10$^{6-9}$M$_\odot$, most black hole growth happens in the AGN 
phase \cite{lynden-bell69,soltan82}. With typical bolometric luminosities $\sim$10$^{45-48}$erg~s$^{-1}$, AGN are amongst 
the most luminous emitters in the Universe, particularly at high energies and radio wavelengths. These luminosities are 
a significant fraction of the Eddington luminosity --- the maximum luminosity for spherical accretion beyond which radiation
pressure prevents further growth --- for a 10$^{8-9}$~M$_\odot$ central black hole. A significant fraction of the total black hole
growth, $\sim$60\% \cite{treister10}, happens in the most luminous AGN (quasars) which are likely triggered by the major merger 
of two massive galaxies, as it is discussed in Sect.\ \ref{trigger}. In the quasar phase, which lasts $\sim$10$^8$ years, the central supermassive
black hole can gain up to $\sim$10$^7$-10$^8$ M$_\odot$, so even the most massive galaxies will have only a few of 
these events over their lifetime. Further black hole growth, mostly in low-luminosity (low Eddington rate) AGN, is 
likely due to stochastic accretion of cold gas, mostly in spiral galaxies \cite{hopkins06a}.

According to the AGN unification paradigm \cite{antonucci93,Urry:95}, a large fraction of these sources, $\sim$75\% locally, 
are heavily obscured by optically and geometrically thick axisymmetric material, which explains many of the observed 
differences among different types of active galaxies. In addition, luminosity \cite{lawrence91} and cosmic epoch
\cite{treister06b} play a significant role. One constraint on the fraction of obscured AGN and its evolution comes from the 
spectral shape of the extragalactic X-ray ``background'' (XRB). Thanks to deep X-ray observations at E$\lsim$10~keV 
performed by {Chandra} and {\it XMM-Newton}, a very large fraction of the X-ray background, $\sim$80\%, has been 
resolved into point sources \cite{hickox06}, the vast majority of them AGN \cite{mushotzky00}. Several studies, the first of 
them $\sim$20 years ago \cite{setti89}, have used a combination of obscured and unobscured AGN to explain the spectral 
shape and normalization of the X-ray background with overall good results. The latest AGN population synthesis 
models \cite{treister09b,draper09}  assume an average ratio of obscured to unobscured AGN of $\sim$3:1 locally, increasing
towards lower luminosities and higher redshifts, as well as a fraction of Compton-thick 
sources (CT; $N_H$$>$10$^{24}$cm$^{-2}$) of $\sim$5-10\%, consistent with the value observed at higher energies,
E=10-100~keV, of $\sim$5\% by {\it INTEGRAL} in the local Universe \cite{treister09b,burlon11}, lower by factors of $\sim$3
than expectations of previous population synthesis models \cite{treister05b,gilli07}. 

Direct black hole mass measurements, either through stellar or gas dynamics, are available only for a few nearby galaxies. 
However, thanks to the tight correlation between the mass of the supermassive black hole and other properties (such as velocity 
dispersion), it has been possible to estimate the black hole mass function at $z$$\simeq$0 
\cite{salucci99,yu02,marconi04,shankar04}. This is commonly done starting from the observed galaxy luminosity or velocity 
function and assuming either a constant black hole to stellar mass ratio \cite{salucci99} or the M-$\sigma$ relation 
\cite{marconi04}.  Both the overall shape of the black hole mass function and the integrated black hole mass density, which 
can only be computed at $z$$\simeq$0, can be used to infer properties of the AGN population. This was first used in the 
so-called ``Soltan's argument'' \cite{soltan82}, which says that the intrinsic bolometric AGN luminosity, $L$, is directly linked to 
the amount of mass accreted by the black hole, $ \dot{M}_{acc}$:

\[
L=\varepsilon \dot{M}_{acc}c^2,
\]

\noindent where $\varepsilon$ is the accretion efficiency and $c$ is the speed of light. A typical value
assumed for the efficiency is $\sim$10\% \cite{soltan82,marconi04}.

Recent comparisons of the black hole mass function to the distribution inferred from the observed AGN luminosity indicate that 
the average efficiency is 8\%, the Eddington ratio is $\sim$50\%, and the average lifetime of the visible AGN phase 
is $\sim$10$^8$ years \cite{marconi04,shankar04}. By studying the black hole mass distribution at the high mass end, 
M$>$10$^9$M$_\odot$, Natarajan \& Treister \cite{natarajan09} found that the observed number of ultra-massive black holes 
is significantly lower than the number density inferred from the AGN hard X-ray luminosity function. They concluded that this 
is evidence for an upper limit to the black hole mass, which can be explained by the presence of a self-regulation 
mechanism.

\begin{figure}[tb]
\begin{center}
\includegraphics[angle=0,scale=0.35]{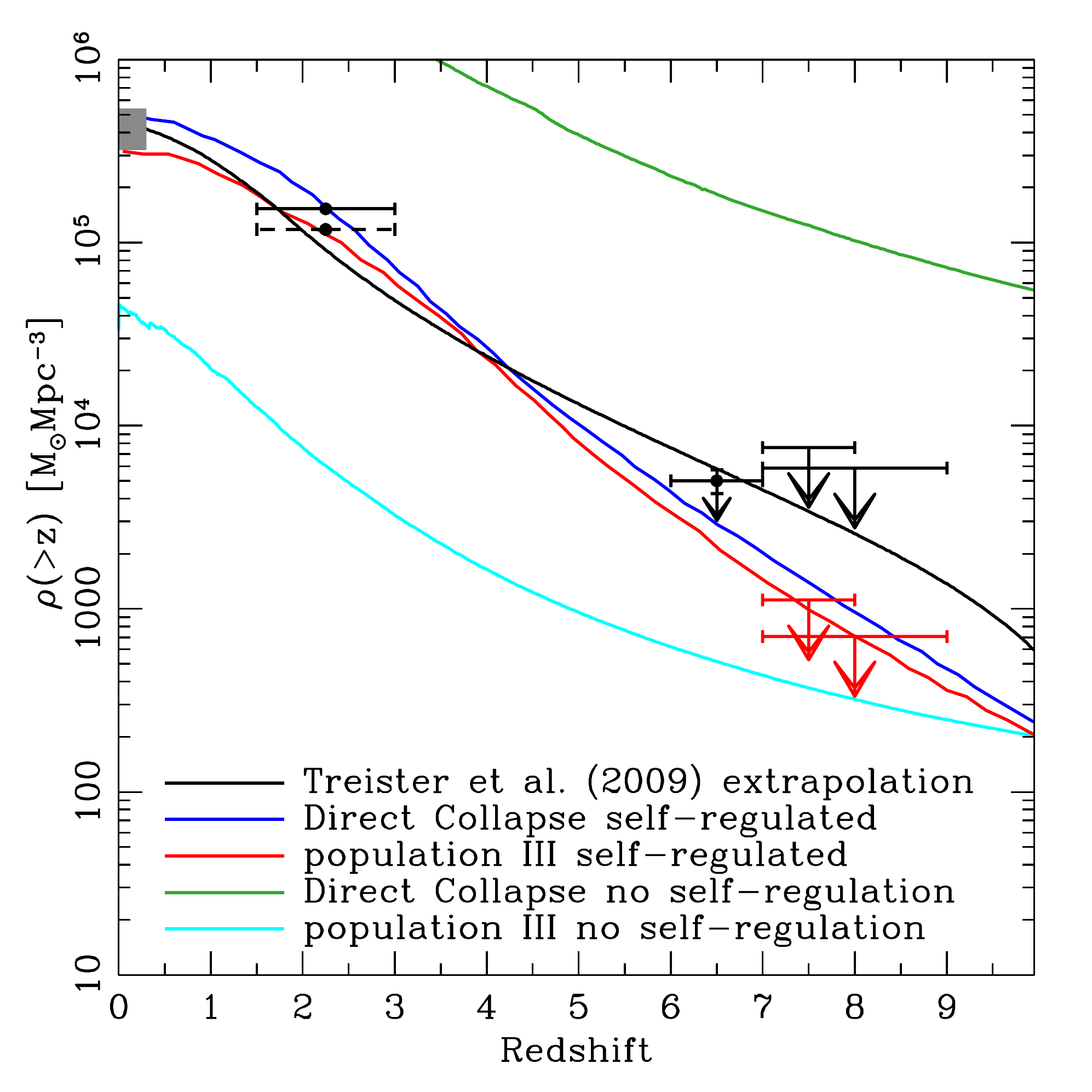}
\end{center}
\caption{
Total accreted mass by supermassive black holes per cubic megaparsec as a function of redshift.
The {\it gray rectangle} shows the  range of values allowed 
by observations of $z\simeq$0 galaxies \cite{shankar09}. 
The data points at $z$$\sim$2 correspond to the values 
obtained from {Chandra} observations of X-ray detected AGN and luminous infrared galaxies \cite{treister10}, while the 
upper limits at $z$$>$6 show the results described in  \cite{treister11} ({\it red} and {\it black} data points from the 
observed-frame soft and hard X-ray band observations respectively). 
The {\it black} solid line shows the evolution of the 
total accreted mass per proper volume element inferred from the extrapolation of AGN luminosity functions 
measured at lower redshifts \cite{treister09b}. 
We over-plot the predictions of black hole and galaxy evolution models \cite{volonteri10a} for non-regulated 
growth of Population-III star remnants ({\it cyan line}) and direct-collapse seeds ({\it green}). The {\it red}  and {\it blue} lines show the predicted BH mass density if self-regulation is incorporated.\label{bhdens_red}
}
\end{figure}

The observed black hole mass density at $z$$\simeq$0 obtained by integrating the black hole mass function, 
ranges from 2.9$\times$10$^5$ \cite{yu02} to 4.6$^{+1.9}_{-1.4}$$\times$10$^5$~M$_\odot$Mpc$^{-3}$ \cite{marconi04};  more recently, Shankar et al. (2009) found 3.2-5.4 $\times$10$^5$~M$_\odot$Mpc$^{-3}$ \cite{shankar09}. For comparison, 
integrating the AGN hard X-ray LF, including the number of Compton-thick AGN constrained by {\it INTEGRAL} and Swift/BAT 
	observations, Treister et al.  obtained a value of 4.5$\times$ 10$^5$ M$_\odot$~Mpc$^{-3}$, perfectly 
consistent with the observed value, indicating that at least locally, X-ray detected AGN can account for most or all of the 
black hole growth \cite{treister09b}. A comparison between the observed black hole mass density at $z$$\simeq$0 and the value derived
from integrating the AGN luminosity function is shown in Figure~\ref{bhdens_red}.

At higher redshifts, the black hole mass density at $z$$\sim$1-3 can be derived from the {Chandra} observations of 
X-ray detected AGN and luminous infrared galaxies at $z$$\sim$2 \cite{treister10}. 
Upper limits to the observed black hole mass density at $z$$>$7 were obtained from X-ray stacking at the position
of high-redshift galaxy candidates in the Chandra Deep Field South by \cite{treister11}. 
Although a strong detection at $z$$\sim$6 was reported \cite{treister11}, 
it was recently questioned by other authors \cite{cowie11,willott11}. In the following, we treat the detection as an upper limit. 
The measurements can be compared to the expectations derived from black hole growth models. 
In particular, we use as a comparison the results in  
\cite{volonteri03}, \cite{volonteri09} and \cite{volonteri10c}. Two ``seed'' formation models are 
considered : those deriving from population-III star remnants (Pop III), and from direct collapse models (D.C.). In this scheme, 
the central black holes accumulate mass via accretion episodes that are triggered by galaxy mergers.  Accretion proceeds in 
one of two modes:  self-regulated or un-regulated.  For each black hole in these models we know its mass at the time when the 
merger starts ($M_{\rm in}$), and we set the final mass through the self-regulated or un-regulated prescription.
These two models differ by the amount of mass a SMBH accretes during a given accretion phase. 
As it can be seen in
Fig.~\ref{bhdens_red}, models that do not incorporate the effects of self-regulation in black hole growth are grossly inconsistent
with the available observational data.

While a clear picture of the history of black hole growth is emerging, significant uncertainties still remain. In particular, 
while the spectral shape and intensity of the extragalactic X-ray background have been used to constrain the AGN 
population, the number of heavily obscured accreting supermassive black holes beyond $z$$\sim$1 is not properly bounded. 
Infrared and deep X-ray selection methods have been useful in that sense, but have not provided a final answer, due to
confusion with star-forming galaxies in the infrared and the effects of obscuration in X-rays. At higher redshifts, the situation 
is even more unclear, and only a few, very rare, high luminosity quasars are known. Unless high-redshift AGN luminosity
functions are pathological, these extreme sources do not represent the typical growing black holes in the early Universe. As 
a consequence, and in spite of recent advances \cite{treister11,natarajan11}, the formation mechanism for the first black 
holes in the Universe is still unknown.

\section{What Triggers Black Hole Growth?}
\label{trigger}

While it is clear now that most galaxies contain a supermassive black hole in their center, 
only a small fraction of these black holes are AGNs.
This indicates that black hole growth is most likely episodic, with each luminous event lasting 
$\sim$10$^7$-10$^8$ years \cite{dimatteo05}. An obvious question is: What triggers these black hole growth episodes?

Major galaxy mergers provide a good explanation, since, as simulations show, they are very efficient in driving gas to the galaxy center 
\cite{barnes91}, where it can be used as fuel for both intense circumnuclear star formation and black hole growth. Indeed, a clear link 
between quasar activity and galaxy mergers has been seen in intensely star-forming galaxies like ultra-luminous
infrared galaxies (ULIRGs) and in some luminous quasars \cite[e.g.,][]{sanders88}. In contrast, many AGN are clearly not in mergers or 
especially rich environments \cite{DeRo:98}. Instead, minor interactions \cite{moore96}, instabilities driven by galaxy bars 
\cite{kormendy04} and other internal galaxy processes might be responsible for these lower activity levels. Understanding the role of 
mergers is further complicated by the difficulty of detecting merger signatures at high redshifts.

In order to reconcile these potentially contradictory observations, it has been suggested that the AGN triggering mechanism
is a function of luminosity and/or redshift \cite[][and others]{finn01}. More recently, \citet{hopkins09a} used five indirect tests 
to conclude that the triggering mechanism is strongly luminosity-dependent and more weakly redshift-dependent, so that only the most 
luminous sources, which are preferentially found at $z$$>$2, are triggered by major mergers \cite{hopkins09a}. Thanks to results from large
AGN surveys, which now include heavily-obscured IR-selected sources, and recent deep high-resolution observations carried out with 
the \emph{Hubble} WFC3 detector, it is now possible to obtain reliable morphological information even for high-$z$, low 
luminosity sources.

\begin{figure}[tb]
\begin{center}
\includegraphics[angle=0,scale=0.35]{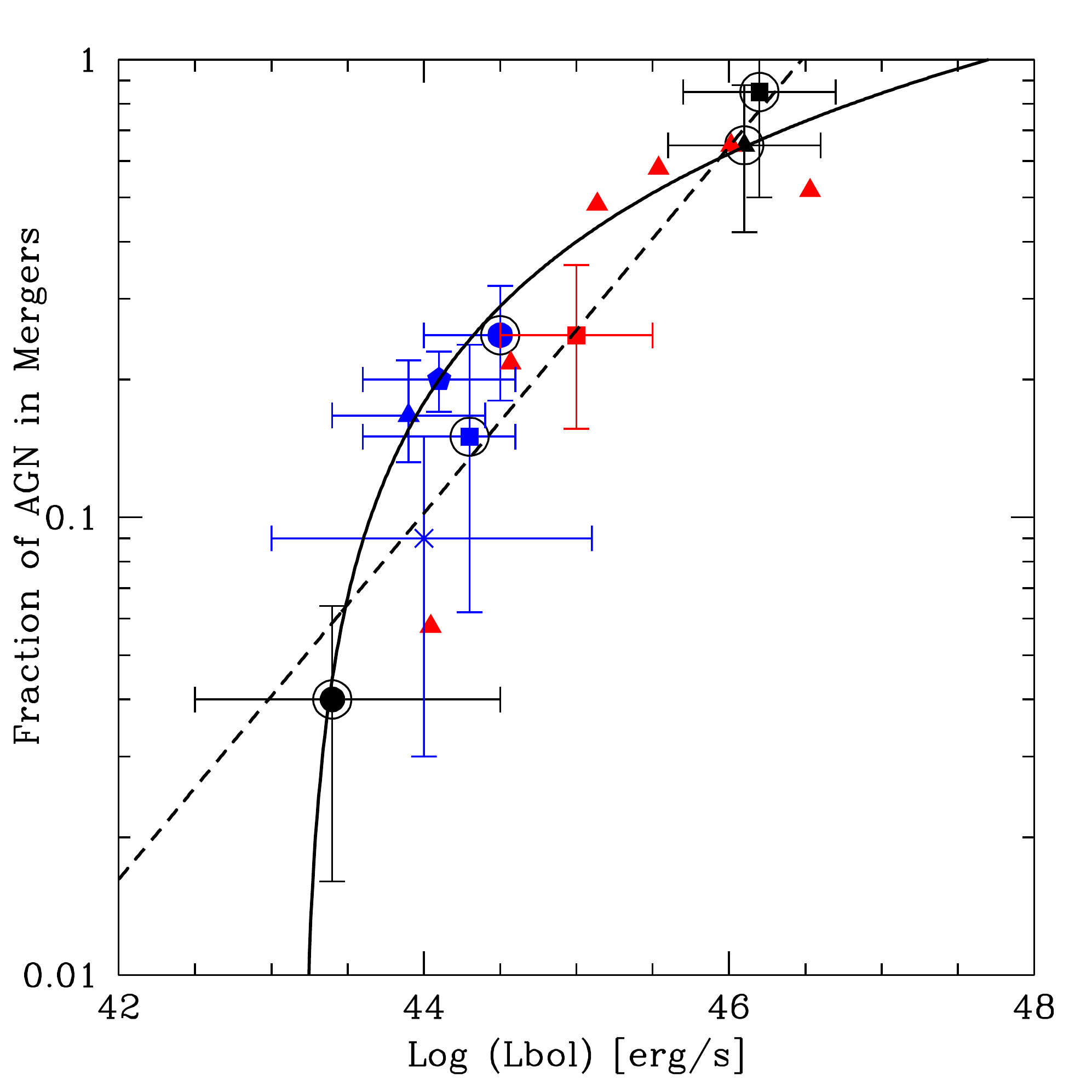}
\end{center}
\caption{Fraction of AGN showing mergers as a function of the AGN bolometric luminosity. Colors indicate AGN selection 
method ({\it red}: infrared, {\it blue}: X-rays, {\it black}: optical) (from \cite{treister12}). 
Encircled symbols show samples at $z$$<$1. {\it Solid line} shows a 
fit to the data assuming a linear dependence of the fraction on $\log$$(L_{bol})$, 
while the {\it dashed line} assumes a power-law dependence.}
\label{merg_frac_lum}
\end{figure}

To measure the fraction of AGN hosted by a galaxy undergoing a major merger as a function of luminosity and redshift, Treister et al. (2012)  
compiled information from AGN samples selected from X-ray, infrared and spectroscopic surveys \cite{treister12}. They studied data from 10 independent 
surveys, which include 874 AGN, spanning a wide range in luminosities, 3$\times$10$^{42}$$<$L$_{bol}$(erg~s$^{-1}$)$<$5$\times$10$^{46}$,
and redshift, 0$<$$z$$<$3. The goal of their work is to determine the physical mechanism(s) that provoked the AGN activity identified in 
these surveys. Only visual morphological classifications have been used, as they are the reliable option to determine if a galaxy is 
experiencing a major merger \cite{darg10}. The fraction of AGN linked to galaxy mergers in these samples has been computed by dividing 
the number of AGN in which the host galaxy has been classified as an ongoing merger or as having major disturbances by the total number 
of AGN. Figure~\ref{merg_frac_lum} shows the fraction of AGN showing mergers as a function of bolometric luminosity, which increases 
rapidly, from $\sim$4\% at 10$^{43}$~erg~s$^{-1}$ to $\sim$90\% at 10$^{46}$~erg~s$^{-1}$.

\begin{figure}[tb]
\begin{center}
\includegraphics[angle=270,scale=0.35]{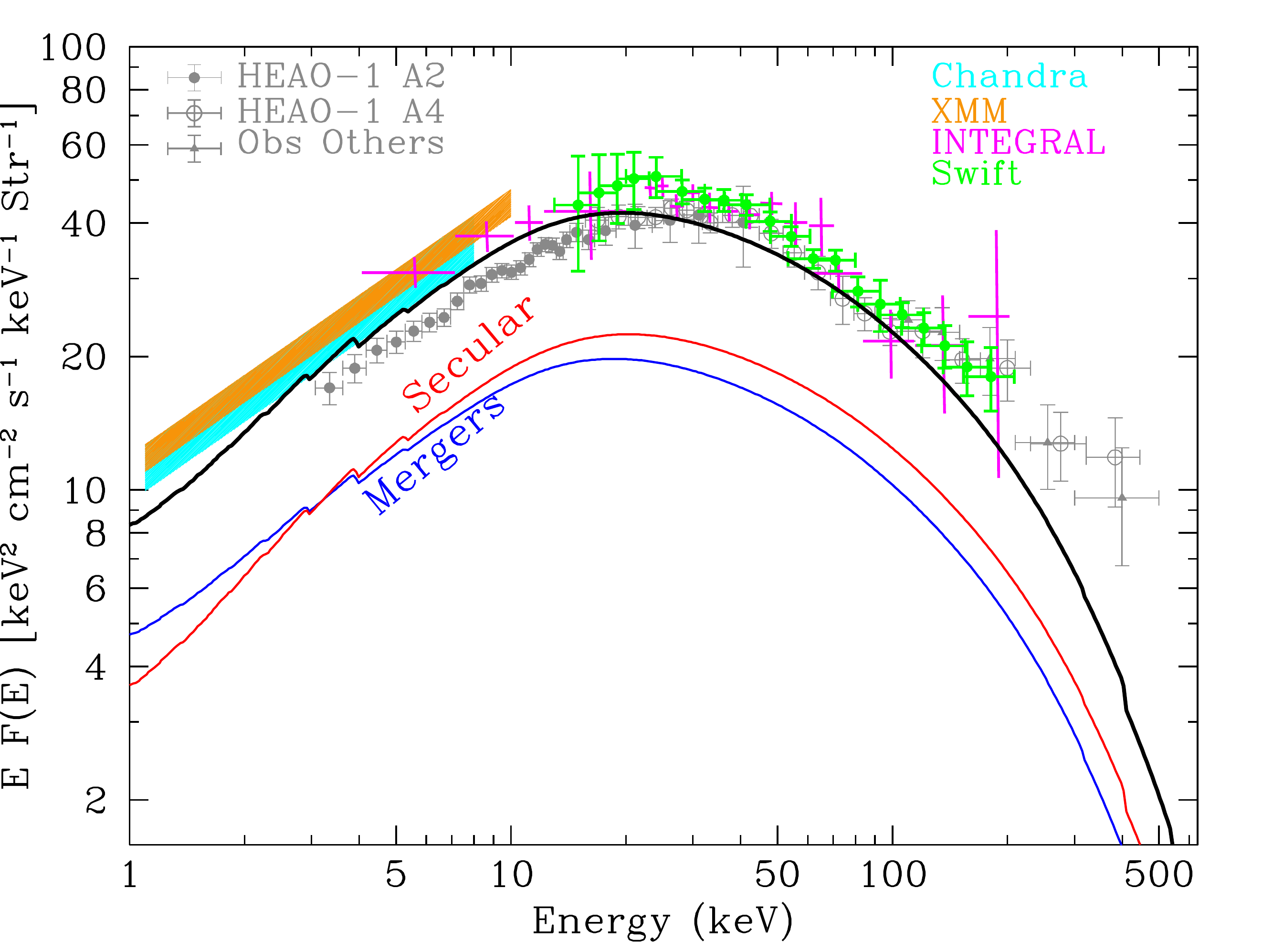}
\end{center}
\caption{Spectral energy distribution of the extragalactic X-ray background, as a function of observed-frame energy. Observational data points are 
summarized in \cite{treister09b}. Merger-triggered AGN ({\it blue line}) contribute roughly equal amounts of light as black hole growth ({\it red line}). Most of the X-ray background emission comes 
from $z$$<$1 \cite{treister09b}, hence the relative importance of secularly-triggered AGN. The extragalactic background light from higher redshift AGN peaks in the optical/UV and is dominated by 
luminous, merger-triggered AGN. The spectral shapes of the merger and secular contributions are slightly different since the fraction of obscured sources is a function
of luminosity.}
\label{xrb_mergers}
\end{figure}

The spectral shape and intensity of the X-ray background can tell us about the average properties of the AGN population. Using the 
models of \cite{treister09b} with the AGN luminosity function of \cite{aird10} and the luminosity dependence of the fraction of AGN 
triggered by major mergers, we can estimate their contribution to the background radiation in X-rays. In Figure~\ref{xrb_mergers} we 
show separately the contributions to the X-ray background from AGN triggered by secular processes and major mergers, which contribute 
nearly equally to the X-ray background. This is because most of the X-ray background comes from $z$$<$1 sources 
\cite[e.g.,][]{treister09b}, where AGN activity due to secular processes is relatively more important. This is particularly true at 
$E$$>$10~keV, where AGN emission is roughly unaffected by obscuration. Because of the luminosity dependence of the
fraction of obscured AGN \cite[e.g.,][]{ueda03}, AGN triggered by secular processes are relatively more obscured than those attributed to
major galaxy mergers, which explains the different spectral shapes in Figure~\ref{xrb_mergers} and the fact that AGN triggered by mergers are more important
at E$<$5 keV. We note that a population of high-luminosity heavily-obscured quasars likely associated with major mergers have been reported by \cite{treister10} and
others. These sources are mostly found at $z$$\sim$2 and show evidence of very high, Compton thick, levels of obscuration. Hence, these sources do not contribute 
significantly to the X-ray background radiation at any energy.

\begin{figure}[tb]
\begin{center}
\includegraphics[width=0.43\textwidth]{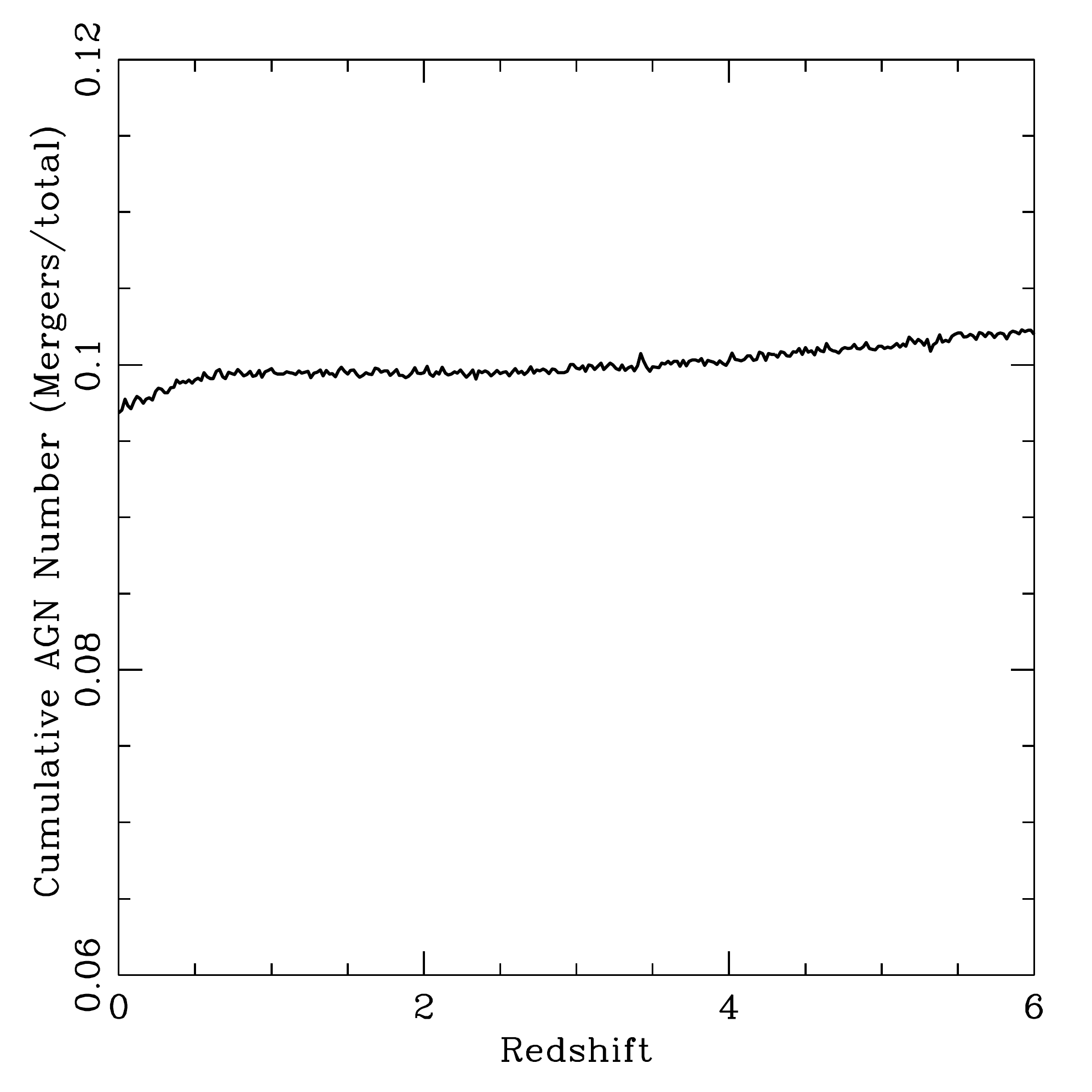}
\includegraphics[width=0.43\textwidth]{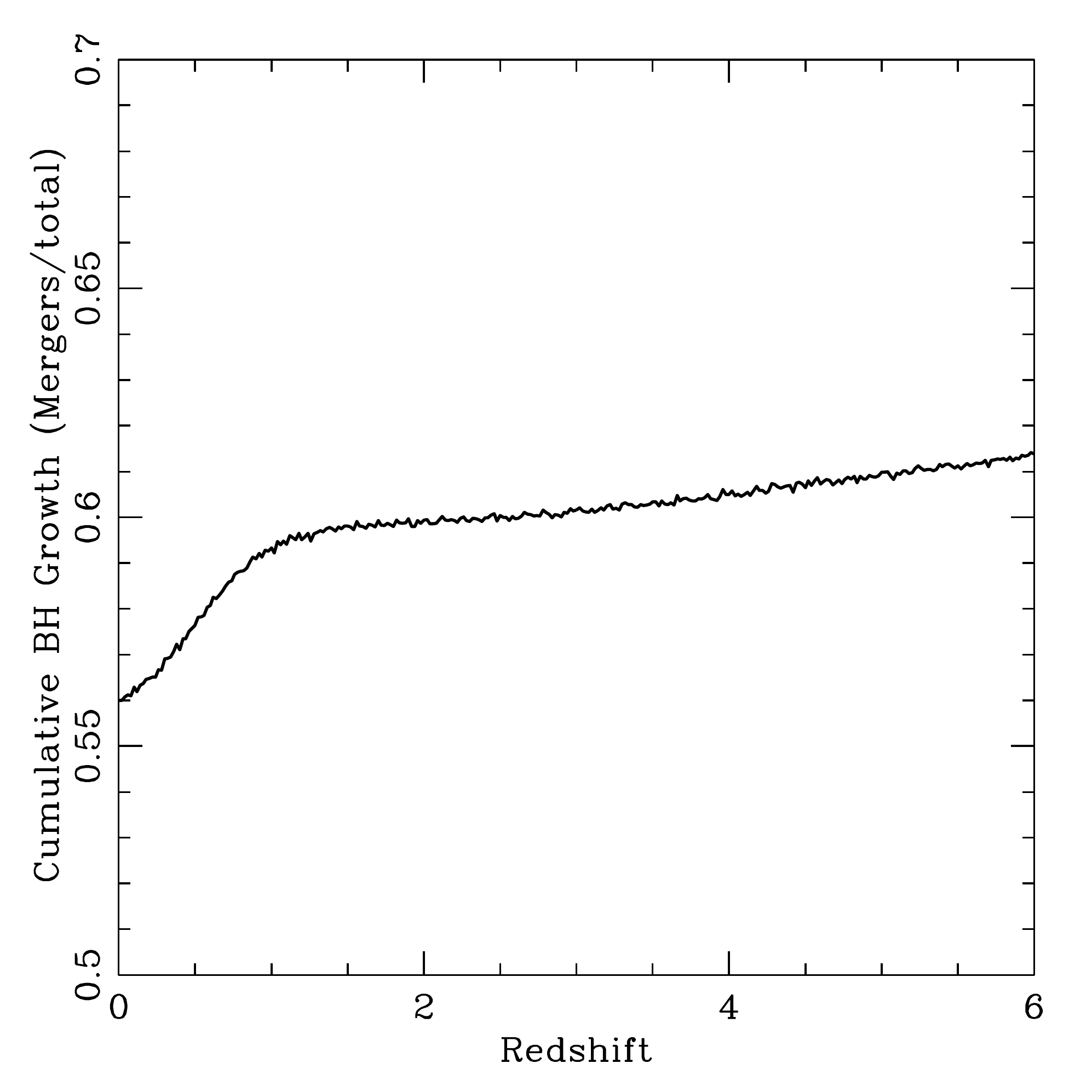}
\end{center}
\caption{{\it Left panel}: Cumulative number of merger-triggered AGN relative to the 
total number of AGN as a function of redshift (from: \cite{treister12}).
While secular-triggered AGN vastly outnumber those triggered by major mergers, by about a factor of $\sim$10, the latter are on average significantly 
more luminous, thus explaining why they constitute $\sim$60\% of black hole accretion. {\it Right panel:} Cumulative fraction of black hole accreted mass in 
AGN triggered by mergers as a function of redshift, assuming a constant efficiency for converting mass to light. Black hole accretion is dominated by merger-triggered AGN at all redshifts but especially at $z$$>$1. At $z$$\sim$1, the much lower gas and merger fractions lead to a dominance of secular processes in AGN accretion.}
\label{red_dep}
\end{figure}

In Figure~\ref{red_dep} we show, as a function of redshift, the amount of black hole growth and number of AGN triggered by major 
galaxy mergers relative to those associated with secular processes. As can be seen and was previously reported 
\cite[e.g.,][]{treister10}, black hole growth occurs mostly in accretion episodes triggered by major galaxy mergers, although secular 
processes are still important. 
This is particularly true at $z$$\gtrsim$2, where there is $\sim$60\% more black hole growth
in merger-triggered AGN than in those growing via secular processes. At lower redshifts, there are relatively fewer galaxy mergers and 
so secular processes become slightly more important. Furthermore, at lower redshifts dry mergers become more common than gas-rich 
major mergers \cite{kauffmann00}. Since the availability of gas is a critical factor in determining the black hole accretion rate, this further 
explains why major mergers are relatively more important at high redshifts. It is interesting to note that the diminishing role of mergers 
coincides with the decline in the space density of black hole growth and with the observed decline in the cosmic star formation rate 
\cite{dahlen07}, i.e., cosmic downsizing. Integrated over the whole cosmic history, to $z$=0, 56\% of the total black hole growth can 
be attributed to major galaxy mergers.

In terms of numbers, the population is strongly dominated by secularly-triggered AGN. Indeed, as can be seen in Figure~\ref{red_dep}, 
$\sim$90\% of AGN at all redshifts are associated with secular processes. This explains the conclusions of previous studies, mostly based on 
X-ray surveys \cite[e.g.][]{cisternas11,schawinski11,kocevski12} of moderate luminosity AGN, which found that normal disk-dominated galaxies 
constitute the majority of the AGN host galaxies. We conclude that while most AGN are triggered by secular processes, most of the black 
hole growth, particularly at high redshifts, can be attributed to intense accretion episodes linked to major galaxy mergers.  
\section{Outlook}
\label{outlook}
The recent progress in our understanding has come from a fleet of telescopes and from the possibility 
to test accretion disk, jet, and particle acceleration models with numerical codes. 
On the observational side, we expect further progress from several observatories; for example: 
\begin{description}
\item [mm and Sub-mm Radio Emission:]
the Atacama Large Millimeter Array (ALMA) will revolutionize our understanding of galaxy evolution. 
Sources of mm and sub-mm emission traced by ALMA include thermal emission of the warm/cold dust, which traces star 
formation, synchrotron radiation associated with relativistic particles and free-free radiation from HII regions. In particular, CO 
rotational transition lines have been used to trace the spatial distribution, kinematics, temperature and mass 
of the molecular gas \cite{yao03}. The  sensitivity of ALMA will allow for the detection of luminous IR galaxies 
($L_{\rm IR}$$>$10$^{11}$$L_\odot$), which have been found to often host a heavily-obscured AGN \cite{treister10b}, up 
to $z$$\sim$10. Furthermore, with ALMA it will be possible to study separately the molecular dust surrounding the central 
black hole and those in star forming regions in the host galaxy. Due to their limited sensitivity and angular 
resolution, currently-available mm/sub-mm telescopes are not ideal to study star-forming regions even in nearby galaxies. 
This will dramatically change thanks to ALMA, which will have orders of magnitude better sensitivity and HST-like angular 
resolution. The first call for ALMA observations was released on March 31, 2011, for observations starting on September 30, 
2011. It is expected that the complete array will be in full operation in 2013. The superb spatial resolution and sensitivity of 
ALMA will allow the identification of the optical/near-IR counterpart of the mm-submm sources.  Furthermore, ALMA will 
directly provide the redshift of the mm-submm sources through the detection of CO rotational transition lines, up to very 
high redshifts. Combining these new data with existing multiwavelength information will finally allow us to complete the
census of supermassive black hole growth since the epoch of cosmic re-ionization.

Using (sub)mm observatories around the world for very long baseline interferometry, it should be possible to 
achieve angular resolutions on the order of 20 $\mu$as \cite{Doel:10}. Such angular resolutions, corresponding to a few gravitational
radii of the SMBHs in the Milky Way and in the nearby radio galaxy M87, may allow us to observe the shadow of 
these two SMBHs. The measurements would provide direct evidence for the presence of an event horizon; 
furthermore, they would allow us to measure the Sgr A* and M87 SMBH spins, and to perform rough tests
of GR in the strong gravity regime.
\item [Hard X-rays:]
launched in June 2012, NuSTAR is the first focusing hard X-ray (5-80 keV) X-ray mission, reaching 
flux limits $\sim$100 times fainter than {\it INTEGRAL} or {\it Swift}/BAT observations and comparable to Chandra 
and {\it XMM-Newton} at lower energies. 
During the first two years of operations, NuSTAR observes, as part of the guaranteed time program, 
two extragalactic fields: the ECDF-S and the central 1 deg$^2$ part of COSMOS, for a total of 
3.1 Msec each. These deep high-energy observations will enable us to obtain a nearly complete AGN survey, including 
heavily-obscured Compton-thick sources, up to $z$$\sim$1.5 \cite{ballantyne11}. A similar mission, ASTRO-H 
\cite{takahashi10}, will be launched by Japan in 2014. Both missions will provide angular resolutions \lsim1$'$, which 
in combination with observations at longer wavelengths will allow for the detection and identification of most growing 
supermassive black holes at $z$$\sim$1. Balloon-borne hard X-ray polarimeters \cite{Beil:11,Pear:11} 
might succeed to measure the polarization of the hard X-ray emission from bright AGN jets. 
The observations could shed light on the magnetic field structure in HSP blazars and could distinguish between
an SSC and EIC origin of the inverse-Compton emission of LSP and ISP sources \cite{Kraw:12}.
\item [$\gamma$-Rays:] the VERITAS array of IACTs has been upgraded in summer 2012 to observe the northern $>100$~GeV $\gamma$-ray sky with a 30\% improved sensitivity. The H.E.S.S.\ collaboration announced first light observations with a large 600 m$^2$ Cherenkov telescope in July 2012 which will enable observations down to energies of 50 GeV. Together with the {Fermi} LAT and MAGIC, these experiments will scrutinize the $\gamma$-ray sky with unprecedented sensitivity. Long-term monitoring programs enabled by {Fermi} LAT's quasi-continuous sky coverage and multiwavelength campaigns with all the $\gamma$-ray telescopes will clarify the existence or non-existence broadband flux and spectral correlations giving us more detailed information about the properties of AGN jets at their bases.
\end{description}
Other experiments, currently in the proposal stage, could revolutionize our understanding of AGNs, including a 
{\it LISA}-type space-based gravitational wave mission (tests of GR, insights into the formation and growth of SMBHs 
from SMBH spin measurements, \cite[e.g.][]{Gair:11}), an {\it IXO}-type high-throughput soft X-ray observatory 
(measurement of the spins 
of a sample of SMBHs through Fe K-$\alpha$ observations) \cite{Barc:10},  a {\it GEMS}-type high-sensitivity X-ray polarimeter 
(geometry of accretion disk coronae and magnetic field structure in jets) \cite{Blac:10}, and the next-generation Cherenkov Telescope Array (time resolved observations of blazars and mapping of the $\gamma$-ray emission from radio galaxies) \cite{Acti:11}.

\section*{Acknowledgements}
HK acknowledges support from NASA (grant NNX10AJ56G), and the Office of High Energy Physics 
of the US Department of Energy.  ET received partial support from Center of Excellence in Astrophysics 
and Associated Technologies (PFB 06) and from FONDECYT  grant 1120061. The authors thank  A.~McCollum 
for help with the references. Special thanks go to Dan Harris and Anca Parvulescu and to the
editors Bing Zhang and Peter Meszaros for carefully reading the manuscript  and providing excellent comments. 

\end{document}